\documentclass[aps,prd,amssymb,final,nobibnotes,nofootinbib,twocolumn,superscriptaddress]{revtex4}
\usepackage{graphicx}
\usepackage{hyperref}
\usepackage{amsmath}
\usepackage{color}
\usepackage{numprint}
\usepackage{float}
\usepackage{ulem}
\usepackage{physics}
\usepackage{mathtools}
\usepackage{dsfont}
\usepackage{subfigure}

\begin{document}
\title{Synchrotron geodesic radiation in Schwarzschild--de Sitter spacetime}

\author{Jo\~ao P. B. Brito}
\email{joao.brito@icen.ufpa.br} 
\affiliation{Faculdade de F\'{\i}sica,
Universidade Federal do Par\'a, 66075-110, Bel\'em, PA, Brazil}

\author{Rafael P. Bernar}
\email{rafael.bernar@icen.ufpa.br} 
\affiliation{Faculdade de F\'{\i}sica,
Universidade Federal do Par\'a, 66075-110, Bel\'em, PA, Brazil}

\author{Lu\'{\i}s C. B. Crispino}
\email{crispino@ufpa.br} 
\affiliation{Faculdade de F\'{\i}sica,
Universidade Federal do Par\'a, 66075-110, Bel\'em, PA, Brazil}

\date{\today}
\begin{abstract}
We analyze the scalar radiation emitted by a source in geodesic circular orbits around a Schwarzschild--de Sitter black hole. We obtain the emitted power using quantum field theory in curved spacetimes framework at tree level. We compare our results with the scalar synchrotron radiation in Schwarzschild spacetime.

\end{abstract}


\maketitle

\section{Introduction}
The recent detection of gravitational waves~\cite{ligo1_2016,ligo2_2016}, emitted by binary black hole systems, and the first visualization of a black hole shadow~\cite{EHT_sombra}, has drawn increasing attention to black hole physics. Moreover, the strong gravity regime, close to these compact objects, plays an important role in general relativity (GR)~\cite{wald_1984} and alternative theories of gravity~\cite{Bertietal2015}, yielding a wealthy scenario for the study of fundamental fields in curved spacetimes, including both their classical and quantum behavior.

On one hand, GR is a very successful classical field theory, both theoretically and experimentally. On the other hand, it is a theory unable to describe the spacetime physics near the singularities appearing in many of its black hole solutions. A quantum theory of gravity is believed to be able to circumvent such problems. There have been several attempts to quantize gravity, with varying degrees of success so far (see the review in Ref.~\cite{Kiefer2005}). In the absence of a final quantum theory of gravity, a more modest approach is the semiclassical framework~\cite{birrel_1982,parker_2009}, in which one considers quantum fields propagating in background spacetimes, which are classical solutions of GR. The quantum field theory (QFT) in curved spacetimes approach has been successful in describing some quantum aspects of gravity, such as the particle creation in dynamic spacetimes~\cite{Parker1969} or near black holes (Hawking radiation)~\cite{hawking_1975}. These findings help in the connection between gravity and quantum theory, leading to important issues such as the black hole information paradox~\cite{Hawking1976} and may even point towards their resolution. QFT in curved spacetimes has provided insights even in flat spacetime, where the Unruh effect, i.e. the fact that accelerated observers in flat spacetime notice the Minkowski vacuum as a thermal bath of particles, is the prime example~\cite{Unruh1976,Crispino2008}. 

The phenomenon of radiation emitted by objects moving along geodesics in a black hole spacetime may be analyzed using the semiclassical framework. The study of such phenomena is important, as black holes found in nature are usually surrounded by accretion disks. This radiation mechanism was originally investigated in Refs.~\cite{misner_1972,misner_et_al_1972}, in which the scalar radiation emitted by sources orbiting a Schwarzschild black hole is studied. The scalar field constitutes a simple model that presents many qualitative results similar to the electromagnetic (vector) and gravitational (tensor) fields. When the source is close to the photon sphere, the radiation is of the synchrotron type, the so-called synchrotron geodesic radiation. Using the QFT in curved spacetimes framework, this type of scalar radiation in asymptotically flat black hole spacetimes was investigated in Refs.~\cite{crispino_2000,castineiras_2007,crispino_2008,crispino_2009,macedo_2012,bernar_2019}, the electromagnetic radiation emission in Ref.~\cite{castineiras_2005} and the gravitational radiation emission in Refs.~\cite{bernar_2017,bernar_2018}. Regarding black holes with nonvanishing cosmological constant, geodesic synchrotron radiation was studied using the Green function framework in Ref.~\cite{cardoso_2002}.

The de Sitter (dS) solution is the simplest solution of GR field equations with a nonvanishing cosmological constant~\cite{desitter_1917_1,desitter_1917_2,hawking_1973,schrodinger}. The study of phenomena in spacetimes asymptotically dS is of great interest, since there is experimental evidence that our Universe is undergoing an accelerated expansion~\cite{riess_1998,perlmutter_1999}. In this more realistic scenario, the black hole solutions are asymptotically dS, rather than asymptotically flat, so that a static chargeless black hole is associated to the Schwarzschild--de Sitter (SdS) spacetime, described by the cosmological constant $\Lambda$, additionally to the geometric mass $M$ of the central Schwarzschild black hole~\cite{kottler_1918,stuchlik_1999,akcay_2011,rindler_2006}.

In this paper, we use QFT in curved spacetime at tree level to investigate the scalar radiation emitted by a source in geodesic circular motion around a SdS black hole. The remaining of this paper is organized as follows. In Sec.~\ref{sec_SdS_black_hole}, we review some features of the SdS spacetime, including the circular geodesic analysis. In Sec.~\ref{sec_field_quantization}, we revisit some aspects of the scalar field theory in this curved background, including the field quantization in the static patch of the SdS spacetime. In Sec.~\ref{sec_scalar_radiation}, using lowest order perturbation theory and numerically obtained solutions for the Klein-Gordon equation, we compute the one-particle-emission amplitude to obtain the power emitted by the source. In the Sec.~\ref{Sec_remarks}, we present our final remarks. We adopt geometrized units in which $c=G=\hbar=1$ and the signature ($-,+,+,+$) for the spacetime metric.

\section{Schwarzschild--de Sitter black holes}
\label{sec_SdS_black_hole}
In this section, we review some important features of the SdS spacetime, which is a spherically symmetric vacuum  solution of GR field equations with a positive cosmological constant $\Lambda>0$ and a black hole with mass $M.$ In static coordinates, the SdS line element  can be written as \cite{akcay_2011,stuchlik_1999,rindler_2006}
\begin{equation}
\label{SdS_line_element}
ds^2 = -f_{\Lambda}(r)dt^2 + \frac{dr^2}{f_{\Lambda}(r)} + r^2(d\theta^2 + \sin^2 \theta d\phi^2),
\end{equation}
with
\begin{equation}
\label{f}
f_{\Lambda}(r) \equiv 1 - \frac{2 M}{r} - \frac{\Lambda}{3}r^2.
\end{equation}
We note that the spacetime described by Eq.~(\ref{f}) has the Killing vectors $\partial_{t},$ associated to translations along $t,$ and $\partial_{\phi},$ as well as $K_1 \equiv \cos \phi \partial_\theta - \cot \theta
\sin \phi \partial_\phi$ and $K_2 \equiv - \sin \phi \partial_\theta - \cot
\theta \cos \phi \partial_\phi,$ associated to rotations on the $2-$sphere.

\begin{figure}
\includegraphics[scale=0.45]{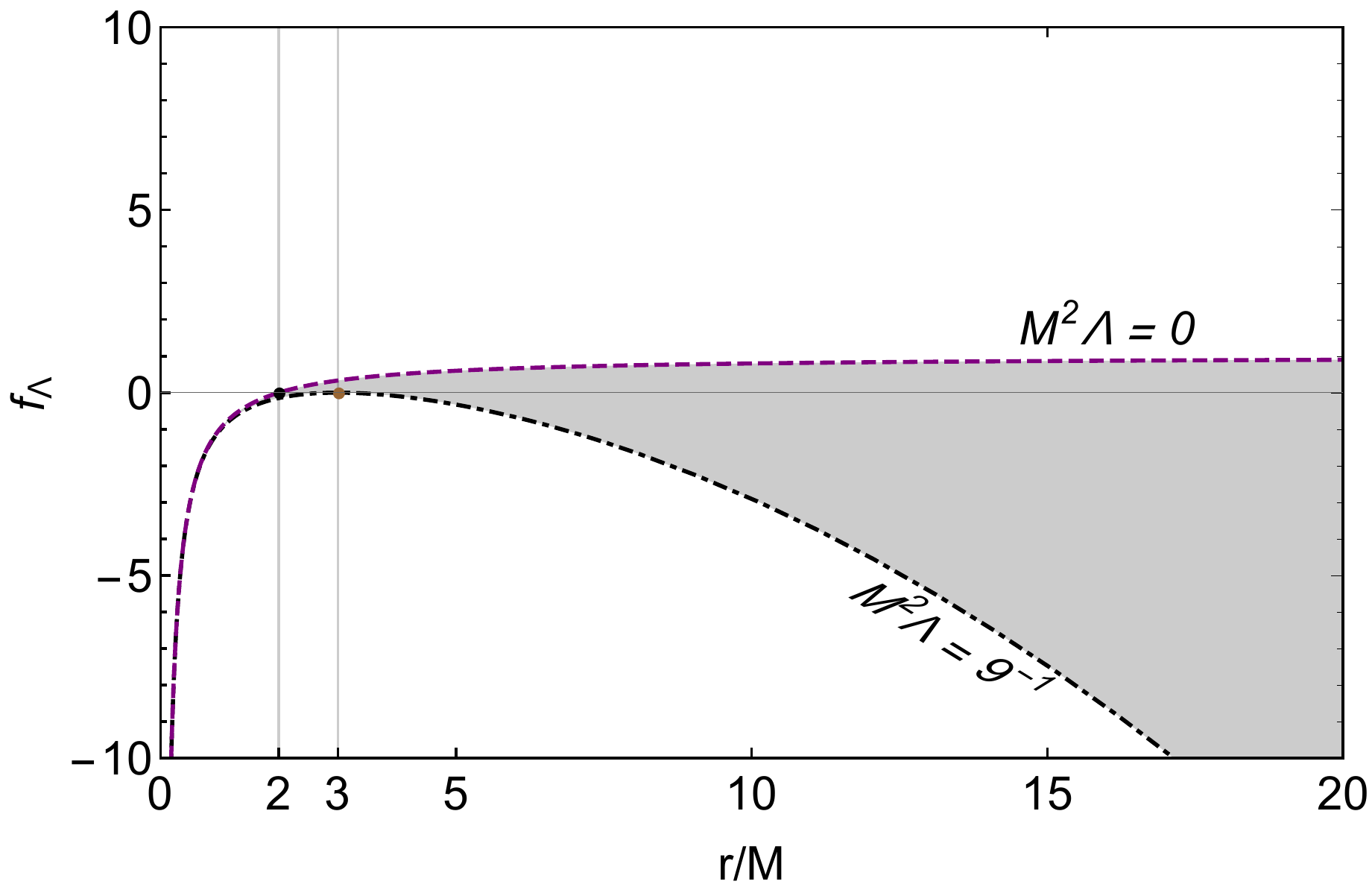}
\includegraphics[scale=0.44]{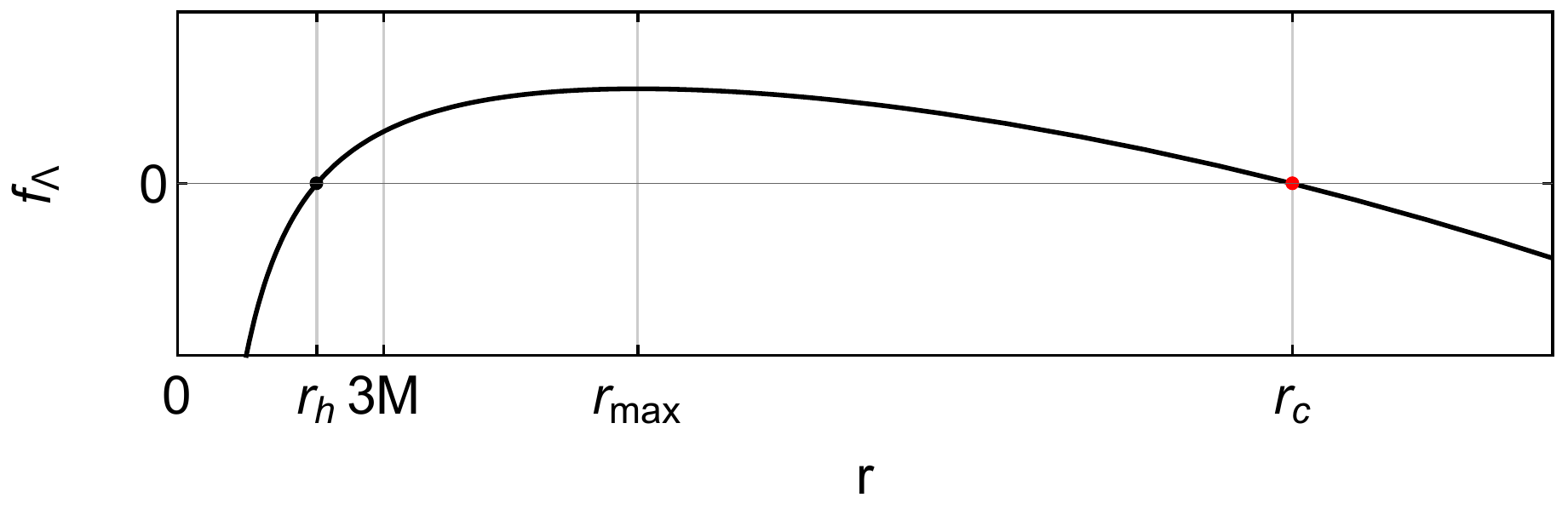}
\caption{Top: The function $f_{\Lambda}(r),$ given by Eq. (\ref{f}), for two different choices of the cosmological constant $\Lambda,$ as indicated. The shaded region encompasses all values of $\Lambda$ in the interval $0 < \Lambda < 1/9M^2.$ Bottom: The function $f_{\Lambda}(r)$ for a given value of $\Lambda>0$, from the interval \eqref{lambda_interval}, with its maximum occurring at the radial position $r_{max}.$}
\label{fr_e_horizons}
\end{figure}

The SdS black hole spacetime presents a cosmological (outer) horizon ($H_c$) and an event (inner) horizon ($H_h$). The radial positions of these hypersurfaces, $r_c$ and $r_h,$ respectively, are obtained by solving
\begin{equation}
\label{f_zero}
f_{\Lambda}(r) = 0.
\end{equation}
For a black hole solution, we must consider the cosmological constant values in the interval,
\begin{equation}
\label{lambda_interval}
 0\leq \Lambda < 1/9M^2.
\end{equation}
In this case, there are up to three real solutions of Eq.~(\ref{f_zero}), two of them are positive (corresponding to the horizons' radial positions, $r_h$ and $r_c$) and one is negative  [$r_-=-(r_h + r_c)$]. We obtain the Schwarzschild solution in the limit $\Lambda \rightarrow 0,$ for which $r_h \rightarrow 2M$ and $r_c \rightarrow + \infty.$ We obtain the dS solution in the limit $M \rightarrow 0,$ for which $r_h \rightarrow 0$ and $r_c \rightarrow \sqrt{3/\Lambda}$ (dS radius).  As the $\Lambda$ term increases from zero (Schwarzschild solution), the two horizons get closer, until they degenerate at the radial position $r_h = r_c = 3M,$ when $\Lambda = \Lambda_{ext} = 1/9M^2$ (extreme case). The behavior of the function $f_{\Lambda}(r)$ is illustrated in Fig.~\ref{fr_e_horizons}. The spacetime is static in the region $r_h < r < r_c$. For $\Lambda > \Lambda_{ext},$ the spacetime is dynamic for all $r>0$~\cite{stuchlik_1999}. 

We shall consider a scalar source rotating around the SdS black hole. In the next subsection, we analyze circular geodesics in the SdS spacetime. 

\subsection{Circular geodesics}
The equations governing the geodesic trajectories in SdS spacetime can be derived from the Lagrangian,
\begin{equation}
\label{lagrangian}
\mathcal{L} = \frac{1}{2}g_{\mu \nu} \dot{x}^{\mu} \dot{x}^{\nu},
\end{equation}
where the metric components $g_{\mu \nu}$ can be obtained from Eq.~(\ref{SdS_line_element}), and the overdot denotes differentiation with respect to an affine parameter (for timelike geodesics, we identify the affine parameter with the free particle's proper time).

The Lagrangian, given by Eq.~(\ref{lagrangian}), is independent of $t$ and $\phi,$ so that we have the following integrals of motion:
\begin{eqnarray}
p_t &=& -\frac{\partial \mathcal{L}}{\partial \dot{t}} = f_{\Lambda}(r) \dot{t} \equiv E, \label{integral_motion_E}\\
p_{\phi} &=& \frac{\partial \mathcal{L}}{\partial \dot{\phi}} = r^2 \dot{\phi} \equiv L. \label{integral_motion_L}
\end{eqnarray}

Without loss of generality, we shall consider the motion in the equatorial plane ($\theta =0$ and $\dot{\theta}=0$). Noting that $2 \mathcal{L} \equiv \epsilon = -1$ $(0)$ for timelike (null) geodesics and using Eqs.~(\ref{integral_motion_E}) and (\ref{integral_motion_L}), we find that the particle motion is entirely described by the energy-balance equation, written as
\begin{equation}
\label{energy}
\dot{r}^2 = E^2 - 2V_{\Lambda}(r),
\end{equation}
with the central potential,
\begin{equation}
\label{central_potenctial}
V_{\Lambda}(r) \equiv \frac{1}{2}f_{\Lambda}(r) \left(-\epsilon + \frac{L^2}{r^2} \right).
\end{equation}
We note that the potential given by Eq.~\eqref{central_potenctial} vanishes at both cosmological and event horizons. For massive particles, there are points of minimum and maximum of the potential~(\ref{central_potenctial}), corresponding to stable and unstable circular orbits, respectively. For massless particles, the potential~(\ref{central_potenctial}) has a maximum at the radial position $r=3M.$ 

For timelike circular orbits, i.e. $\dot{r}=\ddot{r}=0,$ we have the following conserved quantities:
\begin{equation}
\label{circular_E_L}
E^2 = r\frac{f_{\Lambda}(r)^2}{r-3M}, \hspace{1 cm} L^2 = r^2 \frac{M - r^3 \Lambda /3}{r - 3M}.
\end{equation}
Since $E$ and $L$ must be real quantities, circular geodesics exist in the region,
\begin{equation}
\label{circular_range}
3M < r \leq \left( \frac{3M}{\Lambda} \right)^{1/3} \equiv r_{max},
\end{equation}
where $r=r_{max}$ denotes the radial position of the maximum of $f_{\Lambda}(r).$ We note that, for $ 0\ < \Lambda < 1/9M^2$, we have $r_h < r_{max} < r_c$.

\begin{figure}
\includegraphics[scale=0.38]{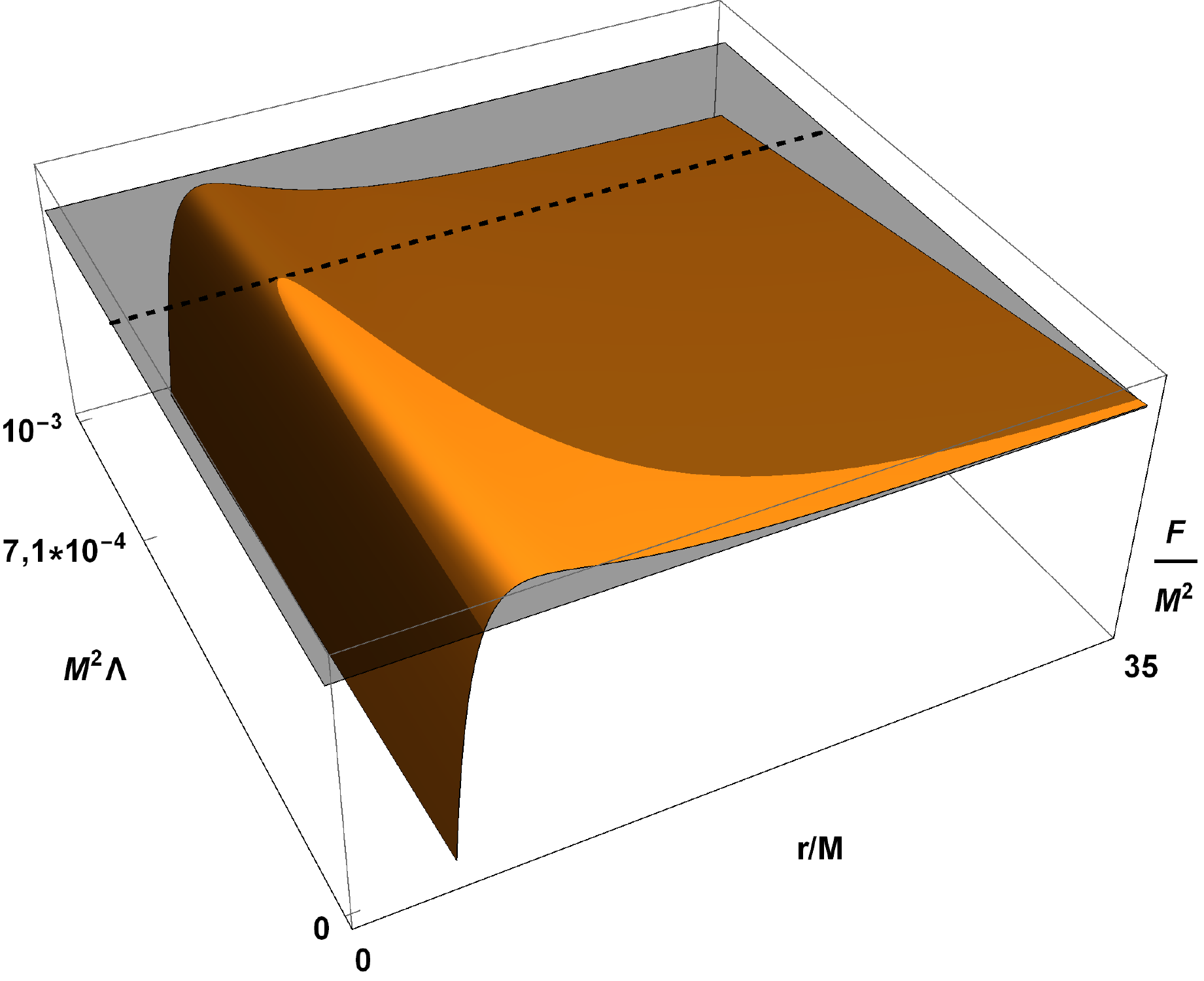}
\caption{The function $F(\Lambda,r),$ given by Eq.~(\ref{stable_condition}). The intersection with the hatched plane marks the zeros of the function.}
\label{F_stability}
\end{figure}

The stability condition for the circular timelike geodesics is obtained considering small radial perturbations on the orbits, as well as by a direct analysis of the potential $V_{\Lambda}(r).$ This condition is found to be~\cite{howes_1979,stuchlik_1999}
\begin{equation}
\label{stable_condition}
F(\Lambda,r) \equiv -4 \Lambda r^4 + 15 \Lambda M r^3 + 3Mr - 18 M^2 \geq 0.
\end{equation}
The function $F(\Lambda,r)$ is illustrated in Fig.~\ref{F_stability}. The points of the surface above the hatched plane select the parameters $\Lambda$ and $r$ for which stable circular orbits can occur. For $M^2\Lambda \leq (64/9)\times10^{-4},$ we have an \textit{innermost stable circular orbit}, at the radial position $r_{isco},$ and an \textit{outermost stable circular orbit}, at the radial position $r_{osco}$ (see, e.g., Ref.~\cite{boonserm_2019}). In the case of $\Lambda =0,$ we have $F(\Lambda,r)\geq 0$ in the interval $6 M \leq r < \infty.$

The orbital angular velocity of the circular timelike geodesics is given by
\begin{equation}
\label{angular_velocity}
\Omega = \frac{d \phi}{dt} = \frac{\dot{\phi}}{\dot{t}} = \sqrt{\frac{M}{r^3} - \frac{\Lambda}{3}},
\end{equation}
which goes to zero as the circular orbit radial position tends to $r_{max},$ defined in Eq.~(\ref{circular_range}). At the radial position $r_{max},$ the gravitational attraction of the central object is balanced by the contribution from the cosmological constant. We can invert Eq. (\ref{angular_velocity}) to obtain $r$ as a function of $\Omega$ and $\Lambda.$
 
Considering $\epsilon = 0$ in Eqs.~(\ref{energy})--(\ref{central_potenctial}), we find that the the radial position $r_0$ of the lightlike circular geodesic is given by the lower limit of the interval in Eq.~(\ref{circular_range}), i.e. $r_0 = 3M.$ Note that $r_0$ is independent of the value of the cosmological constant.

In the next section we analyze the massless scalar field in the SdS background. The field quantization procedure is very similar to that of a Schwarzschild spacetime~\cite{boulware_1975}.

\section{Scalar field quantization}
\label{sec_field_quantization}
\begin{figure}
\includegraphics[scale=0.45]{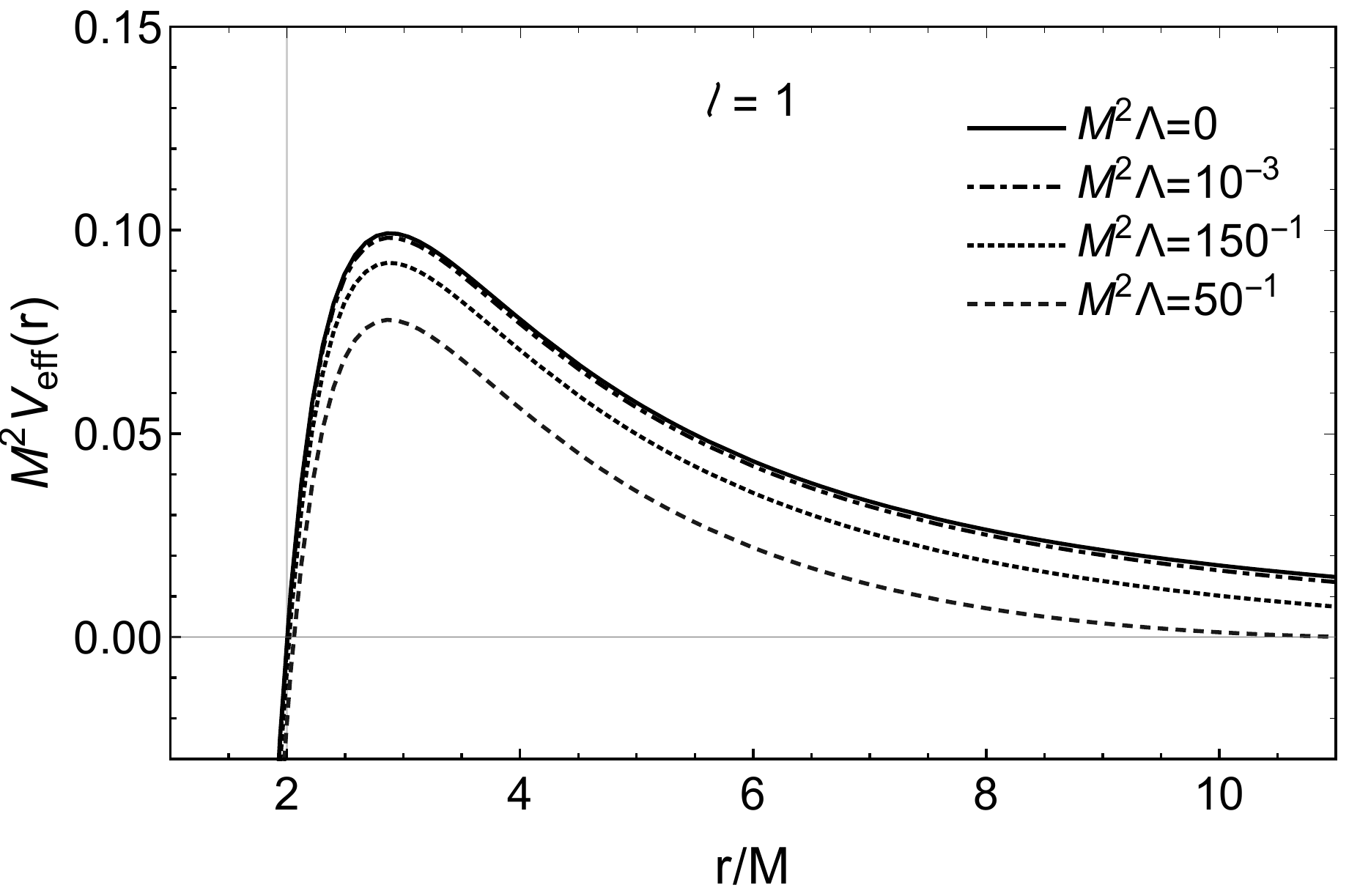}
\caption{The effective potential $V_{eff},$ given by Eq.~(\ref{effective_potential}), with $l=1$ and different choices of the cosmological constant $\Lambda.$}
\label{fig_effective_potential}
\end{figure}

The dynamics of the minimally coupled massless scalar field $\Phi(x)$ in SdS spacetime is governed by the action,
\begin{equation}
\label{action}
S =-\frac{1}{2}\int d^4x \sqrt{-g}\nabla_{\mu} \Phi(x) \nabla^{\mu} \Phi(x),
\end{equation}
from which the equation of motion is obtained to be
\begin{equation}
\label{KG}
\nabla_{\mu}\nabla^{\mu}\Phi(x) = \frac{1}{\sqrt{-g}}\partial_{\mu} \left( \sqrt{-g} g^{\mu \nu} \partial_{\nu} \Phi \right) = 0,
\end{equation}
where $g = - r^4 \sin^2 \theta$ is the determinant of the SdS spacetime metric. The positive-frequency solutions to Eq. (\ref{KG}), with respect to the timelike Killing vector field $\partial_{t},$ can be written in the form, 
\begin{equation}
\label{Mode_positive}
u^k_{\omega l m}(x) = \sqrt{\frac{\omega}{\pi}} \frac{\Psi^k_{\omega l}(r)}{r} Y_{l m}(\theta, \phi) e^{- i \omega t} \hspace{0.4 cm} (\omega > 0),
\end{equation} 
in which $Y_{l m}(\theta, \phi)$ are the scalar spherical harmonics and $\sqrt{\omega / \pi}$ is a normalization constant. The $k$ index in Eq.~(\ref{Mode_positive}) stands for (i) $k=up,$ denoting modes purely incoming from the past event horizon ($\mathcal{H}_h^-$); and (ii) $k=in,$ denoting modes purely incoming from the past cosmological horizon ($\mathcal{H}_c^-$). From Eqs.~(\ref{KG}) and (\ref{Mode_positive}), we find that the function $\Psi^{k}_{\omega l}(r)$ must satisfy the following ordinary differential equation:
\begin{equation}
\label{radial}	
\left( - f_{\Lambda}(r)\frac{d}{dr} \left( f_{\Lambda}(r)\frac{d}{dr}\right) + V_{eff}(r)\right)\Psi^k_{\omega l}(r) = \omega^2 \Psi^k_{\omega l}(r),
\end{equation}
with the effective potential defined by
\begin{equation}
\label{effective_potential}
V_{eff}(r) \equiv f_{\Lambda}(r)\left( \frac{l(l+1)}{r^2} + \frac{2M}{r^3} - \frac{2 \Lambda}{3} \right).
\end{equation}
For $l=0,$ the potential $V_{eff}(r)$ changes sign at $r=r_{max}$ and has a point of minimum in the static region. For $l=1,$ the potential~(\ref{effective_potential}) is illustrated in Fig.~\ref{fig_effective_potential}. We see that as the parameter $\Lambda$ increases, the potential barrier decreases.

Noting that the effective potential (\ref{effective_potential}) vanishes for both $r=r_h$ and $r=r_c,$ we can write the asymptotic solutions of Eq.~(\ref{radial}) in the form,

\begin{equation}
\label{asymptotic_up}
\Psi^{up}_{\omega l} \approx \begin{cases}
A_{\omega l}^{up} \left( e^{i \omega r^*} + \mathcal{R}^{up}_{\omega l}e^{-i\omega r^*} \right), & r \gtrsim r_h, \\
A_{\omega l}^{up} \mathcal{T}^{up}_{\omega l}e^{i\omega r*}, & r \lesssim r_c,
\end{cases}
\end{equation}

\begin{equation}
\label{asymptotic_in}
\Psi^{in}_{\omega l} \approx \begin{cases}
A_{\omega l}^{in} \left( e^{-i \omega r^*} + \mathcal{R}^{in}_{\omega l}e^{i\omega r^*} \right), & r \lesssim r_c, \\
A_{\omega l}^{in} \mathcal{T}^{in}_{\omega l}e^{-i\omega r^*}, & r \gtrsim r_h,
\end{cases}
\end{equation}
where $A_{\omega l}^{k}$ are overall normalization constants to be determined. The tortoise coordinate $r^*$ is implicitly defined by $dr^* \equiv f_{\Lambda}(r)^{-1} dr.$ Hence, $r^*$ goes to $-\infty$ ($+\infty$) in the limit $r \rightarrow r_h$ ($r \rightarrow r_c$). By considering the Wronskian of Eqs.~(\ref{asymptotic_up}) and (\ref{asymptotic_in}), one can show that
\begin{equation}
\label{flux_conservation}
\abs{\mathcal{T}^k_{\omega l}}^2 + \abs{\mathcal{R}^k_{\omega l}}^2 = 1.
\end{equation}

Following the canonical quantization procedure~\cite{birrel_1982,parker_2009,crispino_2000,higuchi_1987,ashtekar_1975}, we may expand the quantum field operator $\hat{\Phi}(x)$ in terms of the creation  ($\hat{a}^{k \dagger}_{\omega l m}$) and annihilation ($\hat{a}^{k}_{\omega l m}$) operators, as
\begin{equation}
\label{field_expansion}
\hat{\Phi}(x) = \sum_{k,l,m} \int_0^{\infty} d\omega \left[u^{k}_{\omega l m}(x)\hat{a}^{k}_{\omega l m} + u^{k *}_{\omega l m}(x) \hat{a}^{k \dagger}_{\omega l m} \right].
\end{equation}
To normalize the modes $u^{k}_{\omega l m},$ we use the Klein-Gordon inner product~\cite{birrel_1982},
\begin{equation}
\left(\Phi, \Psi \right) \equiv i \int_{\Sigma} d\Sigma^{\mu}\left( \Phi^* \left(\nabla_{\mu} \Psi \right) - \Psi \left( \nabla_{\mu} \Phi^* \right) \right),
\label{inner_product}
\end{equation}
in which $d\Sigma^{\mu} = d\Sigma n^{\mu},$ with $n^{\mu}$ being a future directed unit vector orthogonal to the Cauchy surface $\Sigma$ (e.g., the $t=constant$ hypersurface $\Sigma_t$). Since $\hat{\Phi}$ and $\hat{\Psi}$ satisfy Eq. (\ref{KG}), one can show that the inner product (\ref{inner_product}) is independent of the choice of the hypersurface $\Sigma$~\cite{hawking_1973,parker_2009}. By requiring the orthogonality conditions,
\begin{equation}
\label{u_orthogonality}
\left(u^{k}_{\omega l m},u^{k'}_{\omega' l' m'} \right) = \delta_{k k'} \delta_{l l'} \delta_{m m'} \delta(\omega - \omega')
\end{equation}
and
\begin{equation}
\label{u_ortogonality_null}
\left( u^k_{\omega l m},u^{k'*}_{\omega' l' m'} \right) = \left( u^{k*}_{\omega l m},u^{k'}_{\omega' l' m'} \right) = 0,
\end{equation}
one can show that the creation and annihilation operators satisfy the usual nonvanishing commutation relations,
\begin{equation}
\label{comutation_relation}
\left[\hat{a}^k_{\omega l m},\hat{a}^{k \dagger}_{\omega l m} \right] =  \delta_{k k'} \delta_{l l'} \delta_{m m'} \delta(\omega - \omega').
\end{equation}

The vacuum state is defined as the quantum state annihilated by all $\hat{a}^k_{\omega l m}$~\cite{fulling_1973},
\begin{equation}
\label{vacuum}
\hat{a}^{k}_{\omega l m} \ket{0} \equiv 0, \hspace{1 cm} \forall \hspace{0.3 cm} (k, \omega, l, m),
\end{equation}
and the one-particle-state is constructed as
\begin{equation}
\label{estad_uma_partic}
\hat{a}^{k \dagger}_{\omega l m} \ket{0} = \ket{k; \omega l m}.
\end{equation}

Using Eqs.~(\ref{inner_product})--(\ref{u_ortogonality_null}) and the differential equation for $\Psi^k_{\omega l}$ written in terms of the tortoise coordinate, we can readily obtain (up to a phase) the overall normalization constants of Eqs. (\ref{asymptotic_up}) and (\ref{asymptotic_in}), namely 
\begin{equation}
\label{normalization_constant}
A^{up}_{\omega l m} = A^{in}_{\omega l m} = \frac{1}{2 \omega}.
\end{equation}

In the next section, we consider the scalar field coupled to a classical matter source in a SdS spacetime, performing a geodesic circular orbit around the black hole.

\section{Scalar radiation and emitted power}
\label{sec_scalar_radiation}
We consider the scalar source moving along an equatorial circular (timelike) geodesic at $r=R,$ with constant angular velocity $\Omega(R),$ given by Eq.~(\ref{angular_velocity}). The source is described by
\begin{equation}
\label{current}
j(x) = \frac{\sigma}{\sqrt{-g} u^0} \delta(r-R) \delta(\theta - \pi/2)\delta(\phi - \Omega t),
\end{equation}
such that $\int d\beta^{(3)}j(x) = \sigma,$ where $\beta^{(3)}$ is a hypersurface orthogonal to the particle's $4-$velocity. The constant $\sigma$ determines the magnitude of the source-field interaction. The particle's $4-$velocity is given by
\begin{equation}
\label{four_velocity}
u^{\mu}(R) = \gamma \left(1,0,0,\Omega \right),
\end{equation}
with the normalization factor,
\begin{equation}
\label{gamma_factor}
\gamma = \frac{1}{(f_{\Lambda}(R) - R^2 \Omega^2)^{1/2}}.
\end{equation}

The source-field coupling is described by the following interaction action:
\begin{equation}
\label{interaction_action}
\hat{S}_{I} = \int d^4x \sqrt{-g} j(x) \hat{\Phi}(x),
\end{equation} 
where $\sigma$ can be regarded as a coupling constant that determines the magnitude of the interaction between the field and the source.

Due to the interaction between the field and the source, there exists a nonvanishing probability for the radiation of scalar quanta. To lowest order in perturbation theory, the transition amplitude from the vacuum-state, defined in Eq.~(\ref{vacuum}), to the one-particle-state, with quantum numbers $k,$ $l,$ $m$ and energy $\omega,$ is given by~\cite{itzykson_1980}
\begin{equation}
\label{probability_amplitude}
A^{k; \omega l m}_{em} = \bra{k; \omega l m} i \hat{S}_I \ket{0} = i \int d^4x \sqrt{-g} j(x)u^{k *}_{\omega l m}(x).
\end{equation}
It follows that the probability amplitude, given by Eq.~(\ref{probability_amplitude}), is proportional to $\delta(\omega - m\Omega),$ i.e. there is only emission of scalar particles with $\omega_m \equiv m\Omega.$ Since $\omega_m$ and $\Omega$ are positive quantities, we have that $m \geq 1.$

The emitted power (with fixed $k,$ $l$ and $m$) is
\begin{equation}
\label{partial_power_implicit}
W^{k; l m}_{em} = \int_{0}^{\infty} d\omega \omega \frac{\abs{A^{k; \omega l m}_{em}}^2}{T}.
\end{equation} 
Assuming that the source radiates during the whole range of coordinate time $t,$ with $-\infty < t < \infty,$ we can write $T=\int dt = 2 \pi \delta(0)$ \cite{breuer_1975,crispino_1998}.

The emitted power (\ref{partial_power_implicit}) is found to be
\begin{equation}
\label{partial_power}
W^{k;l m}_{em} = 2 \sigma^2 \omega_m^2 \left( f_{\Lambda}(R) - R^2 \Omega^2 \right) \abs{\frac{\Psi^k_{\omega_m l}}{R}}^2 \abs{Y_{l m}\left(\frac{\pi}{2}, \Omega t \right)}^2,
\end{equation}
where the total power is obtained by summing over $k$ ($in$ and $up$), $l \geq 1$ and $1 \leq m \leq l,$ namely
\begin{equation}
\label{total_power}
W_{em} = \sum_{k=in}^{up} \sum_{l=1}^{\infty} \sum_{m=1}^{l} W^{k;l m}_{em}.
\end{equation}
We note that there is no emission for odd values of $l+m,$ since the time independent quantity $\abs{Y_{l m}(\pi/2, \Omega t)}^2$ vanishes in these cases. For even values of $l+m,$ we have~\cite{gradshteyn_1980}
\begin{equation}
\label{harmonic_even_values}
\abs{Y_{l m}(\pi/2, \Omega t)}^2 = \frac{2l+1}{4\pi}\frac{(l+ m - 1)!!(l - m - 1)!!}{(l+m)!!(l-m)!!}.
\end{equation}

For the computation of the (total) emitted power, we obtained the quantity $\abs{\Psi^k_{\omega_m l}}$ by solving Eq.~(\ref{radial}) numerically~\cite{bernar_2017}. In the next section, we present a selection of our results.

\section{Results}
\label{sec_results}
We numerically integrate Eq.~(\ref{radial}) for each $k = in$ and $k = up$ modes. The boundary conditions to be satisfied are given by Eqs.~(\ref{asymptotic_up}) and (\ref{asymptotic_in}), with suitable values of $r.$ We choose $r/M \geq r_h/M + \delta$ for $r$ values near the event horizon, and $r \leq r_c/M - \delta,$ for $r$ values near the cosmological horizon, with
\begin{equation}
\label{erro_num}
\delta = 10^{-5}.
\end{equation}
The numerical error is related to the magnitude of $\delta$.

In order to obtain the transmission and reflection coefficients, $\abs{\mathcal{T}^k_{\omega l}}^2$ and $\abs{\mathcal{R}^k_{\omega l}}^2,$ respectively, we compare the solutions obtained numerically for $\Psi^k_{\omega l}$ and $\frac{d}{dr}\left( \Psi^k_{\omega l} \right),$ with the asymptotic solutions expressed by Eqs.~(\ref{asymptotic_up}) and (\ref{asymptotic_in}), requiring the usual probability flux conservation, given by Eq.~(\ref{flux_conservation}), to be satisfied. 

As an estimation of the numerical error, we may define the quantity,
\begin{equation}
\label{1TR}
\mathrm{Err}^{k}_{\omega l} \equiv \abs{\mathcal{T}^{k}_{\omega l}}^2 + \abs{\mathcal{R}^{k}_{\omega l}}^2 - 1,
\end{equation}
which was kept as $\mathrm{Err}^{k}_{\omega l} \ll 1$.

\begin{figure}
\includegraphics[scale=0.33]{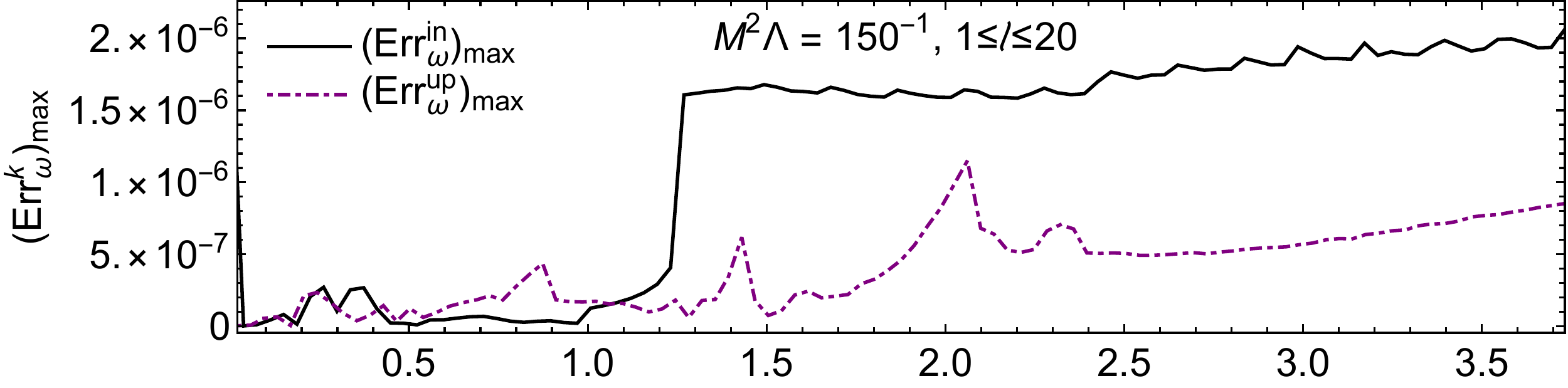}
\includegraphics[scale=0.33]{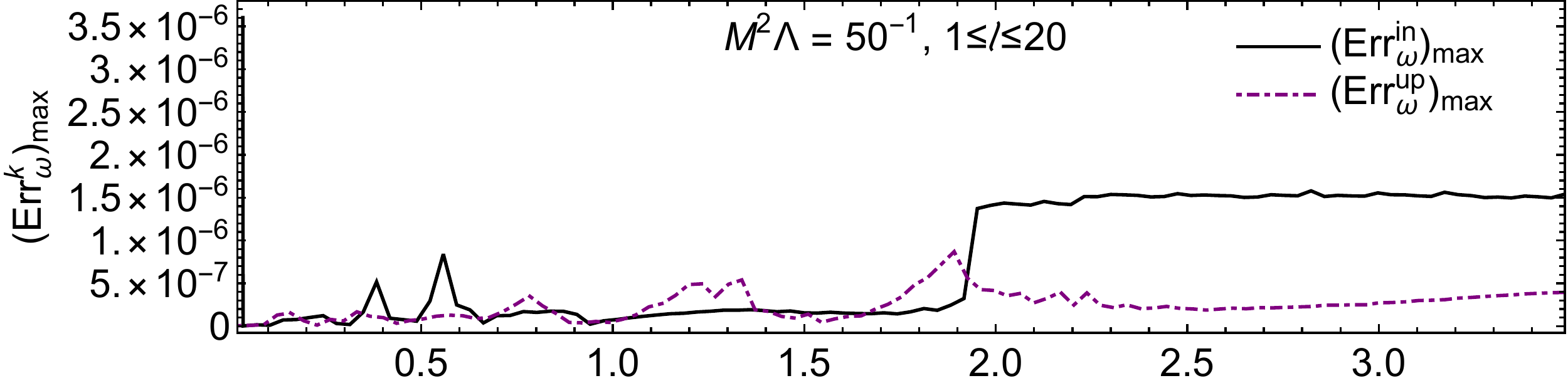}
\includegraphics[scale=0.33]{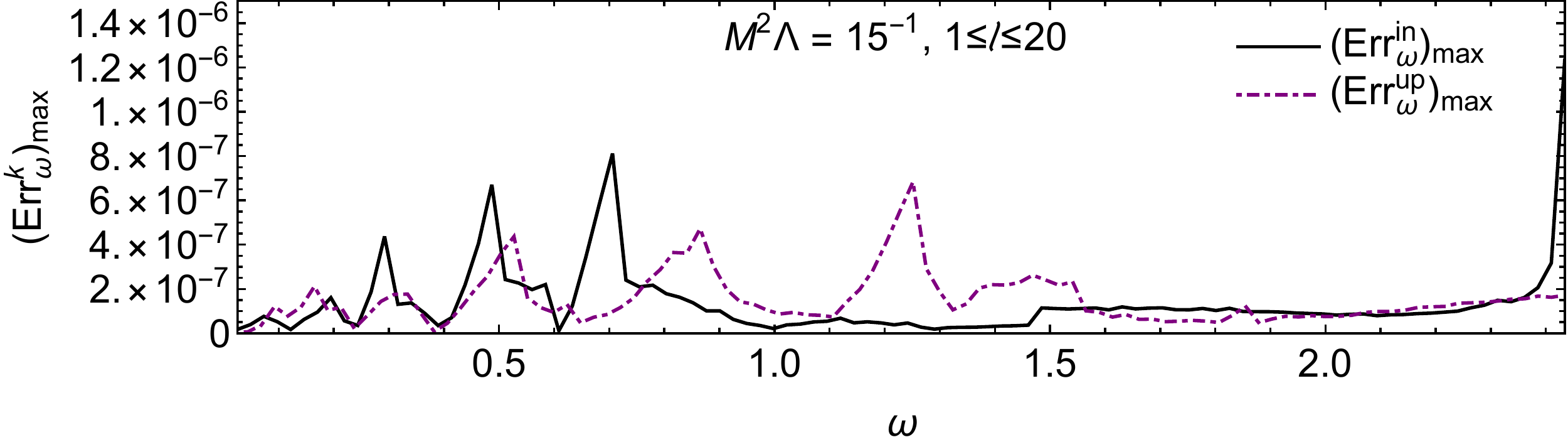}
\caption{The quantity $\left(\mathrm{Err}^{k}_{\omega}\right)_{max},$ as a function of $\omega$, for $k=up$ and $k=in$ with $M^2 \Lambda = 150^{-1}$ (top); $M^2 \Lambda = 50^{-1}$ (middle); and $M^2 \Lambda = 15^{-1}$ (bottom). We consider the interval $ 1 \leq l \leq 20$.}
\label{error}
\end{figure}

\begin{figure}
\includegraphics[scale=0.33]{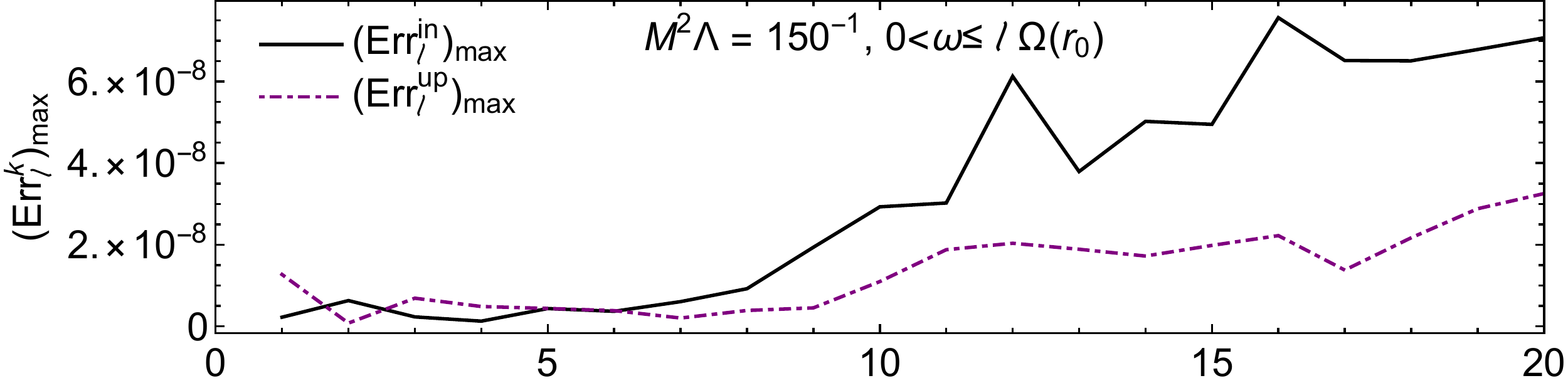}
\includegraphics[scale=0.33]{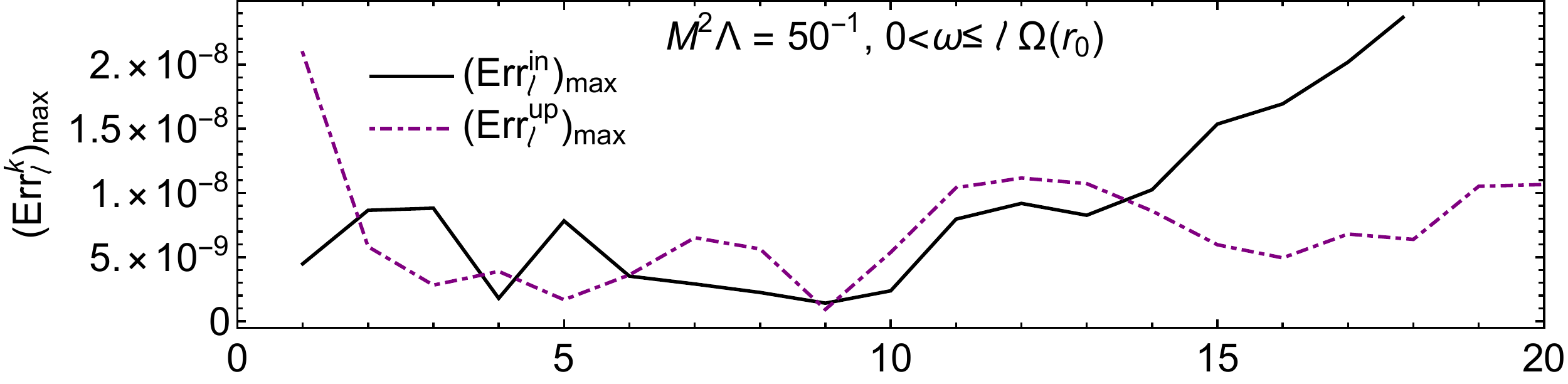}
\includegraphics[scale=0.33]{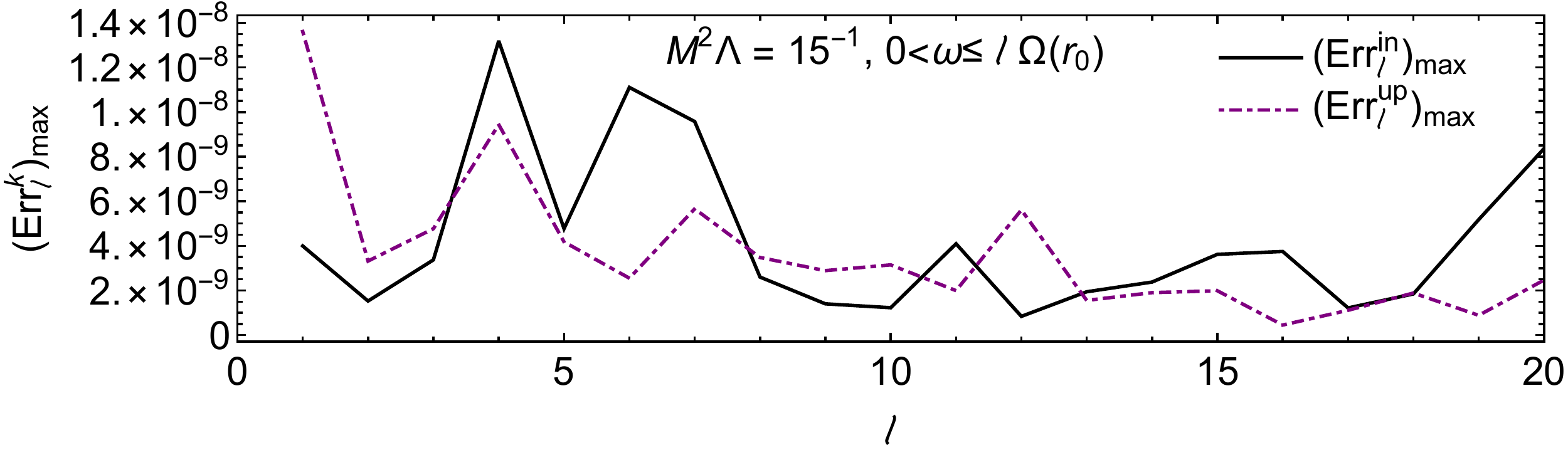}
\caption{The quantity $\left(\mathrm{Err}^{k}_{l}\right)_{max},$ as a function of $l$, for $k=up$ and $k=in$ with $M^2 \Lambda = 150^{-1}$ (top); $M^2 \Lambda = 50^{-1}$ (middle); and $M^2 \Lambda = 15^{-1}$ (bottom). We consider the interval $ 0 < \omega \leq l \Omega(r_0)$.}
\label{error_l}
\end{figure}

To exemplify the numerical errors, in Fig.~\ref{error} we plot $\left(\mathrm{Err}^{k}_{\omega}\right)_{max},$ i.e. the maximum value of the error $\mathrm{Err}^{k}_{\omega l},$ 
considering all values of $l$ in the interval $ 1 \leq l \leq 20,$ for a given $\Lambda$, as a function of $\omega$.
We see that this maximum error is of the order $10^{-6}$, for all values of $\omega.$
Analogously, in Fig.~\ref{error_l}, we plot $\left(\mathrm{Err}^{k}_{l}\right)_{max},$ i.e. the maximum value of the error $\mathrm{Err}^{k}_{\omega l},$ 
considering all values of $\omega$ in the interval $ 0 < \omega \leq l \Omega(r_0),$ for a given $\Lambda$, as a function of $l$.

In general, the emitted power of radiation starts increasing from zero, at $R=r_{max},$ reaches a maximum and then decreases to zero again as $R \rightarrow 3M.$ The radial position of the peak of emission approaches $r=3M$ as we increase the multipole number $l.$ The emitted power associated to $in$ modes is generally dominant, except for the region close to $r=3M,$ in which the $up$ modes start to give a significant contribution. We note that, for orbits close to $r=3M,$ the major contribution ($>97 \%$) to the emitted power, for a given multipole number $l,$ comes from the $l=m$ mode, similarly to what happens in asymptotically flat spacetimes~\cite{Ruffini1972,breuer_1973}.

\begin{figure}[h!]
\center
\includegraphics[scale=0.45]{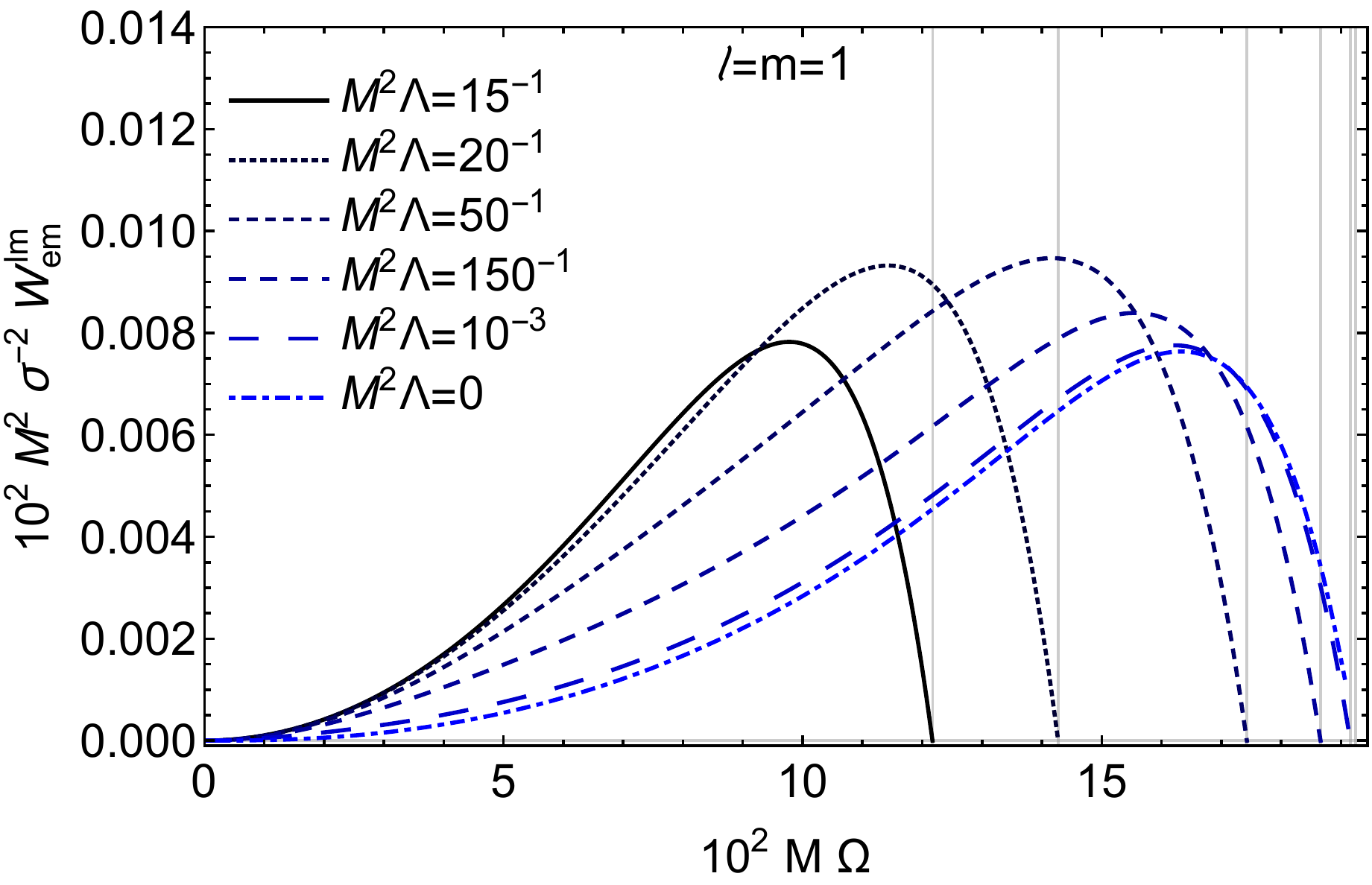}
\includegraphics[scale=0.45]{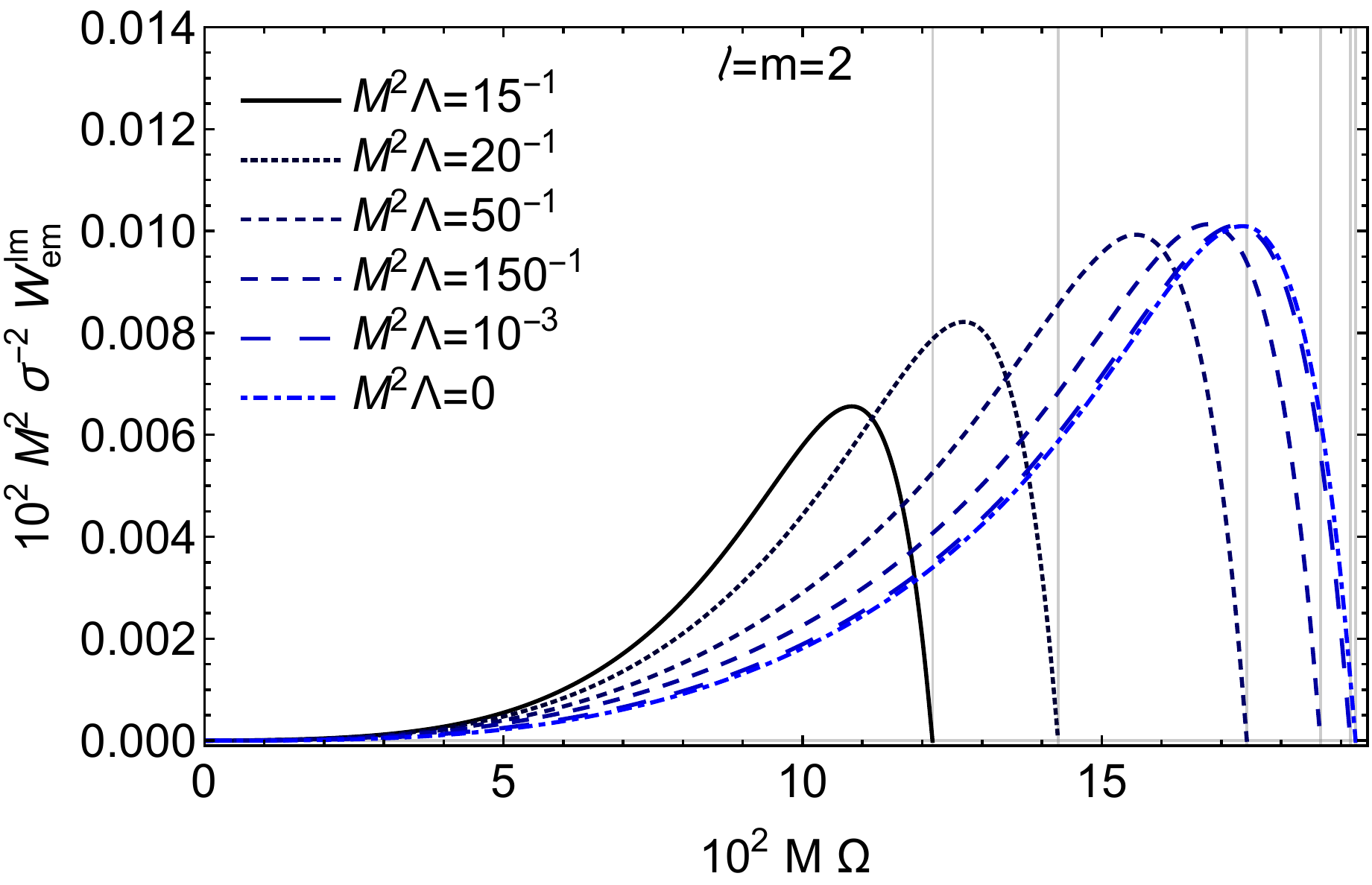}
\includegraphics[scale=0.45]{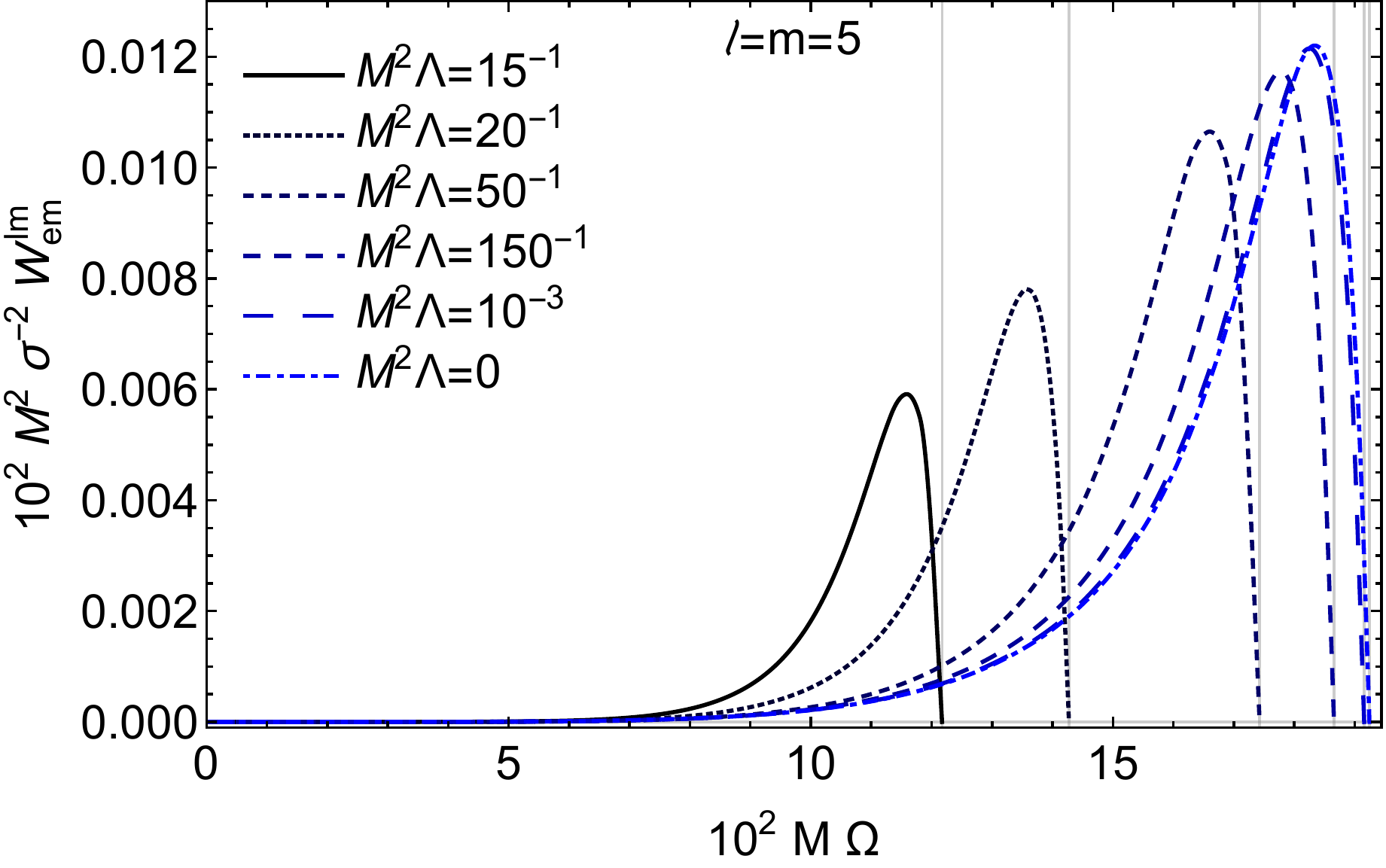}
\caption{The emitted power as a function of $\Omega,$ given by the sum of the $in$ and $up$ modes, for $l=m=1$ (top), $l=m=2$ (middle) and $l=m=5$ (bottom), with different choices of the parameter $\Lambda.$ The curves are plotted up to the values of $M \Omega$ corresponding to the radial position $R=3M$ (indicated by the vertical lines).}
\label{Pot_Tot_various_lamb}
\end{figure}

 In Fig.~\ref{Pot_Tot_various_lamb}, we plot the emitted power by the orbiting source as a function of $\Omega,$ for a fixed value of $l=m,$ for different choices of $\Lambda.$ We see that, for $l=m=1,$ the peak of emission starts increasing with $\Lambda$, but after a certain value of the cosmological constant ($\Lambda \sim 30^{-1}M^{-2}$) the peak starts to decrease. This behavior changes  for higher values of $l=m$, with the peak of emission monotonically decreasing, as the value of $\Lambda$ is increased. 

\begin{figure*}
\center
\includegraphics[scale=0.45]{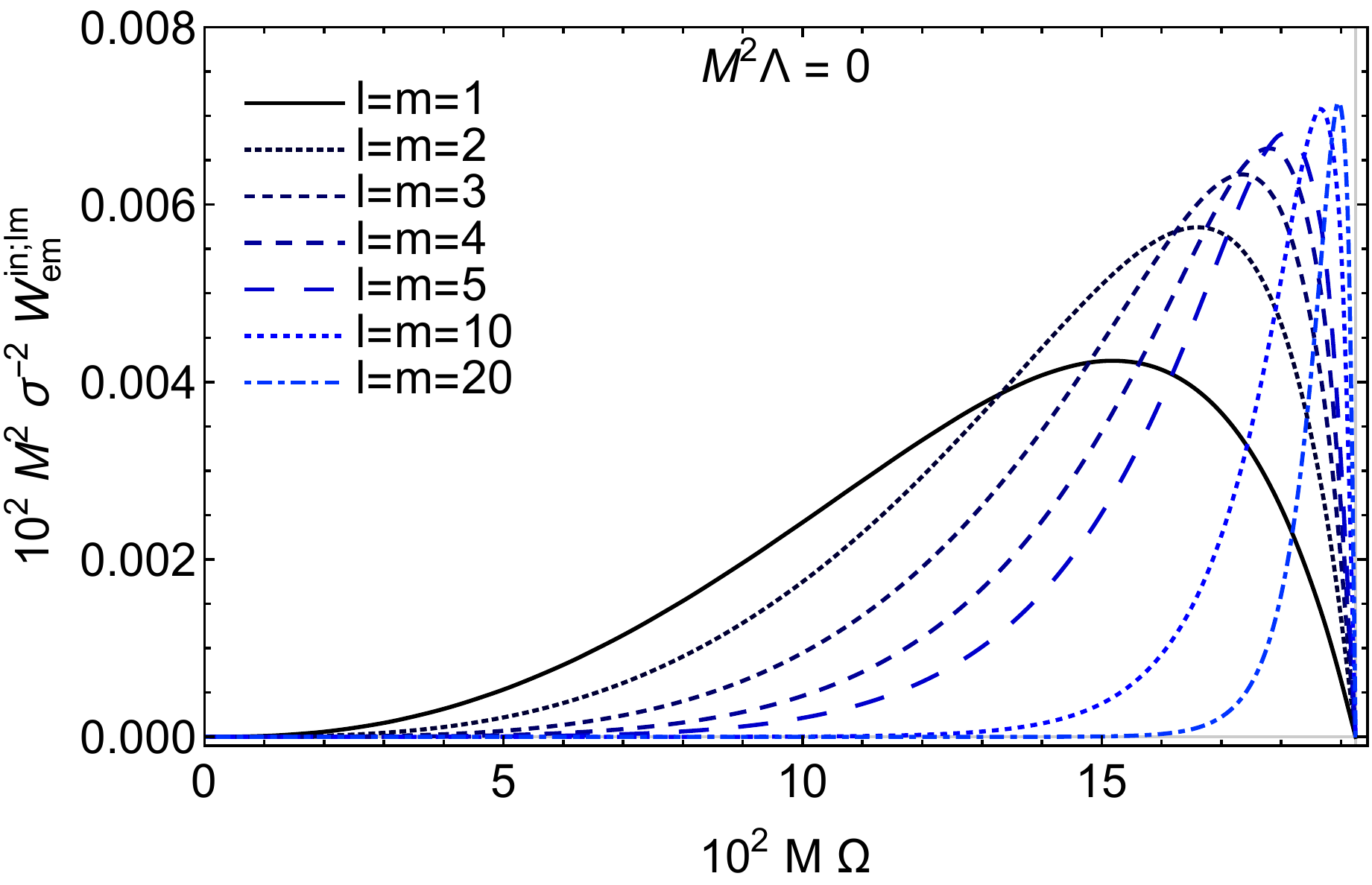} \hspace{0.2 cm}\vspace{0.3 cm}
\includegraphics[scale=0.45]{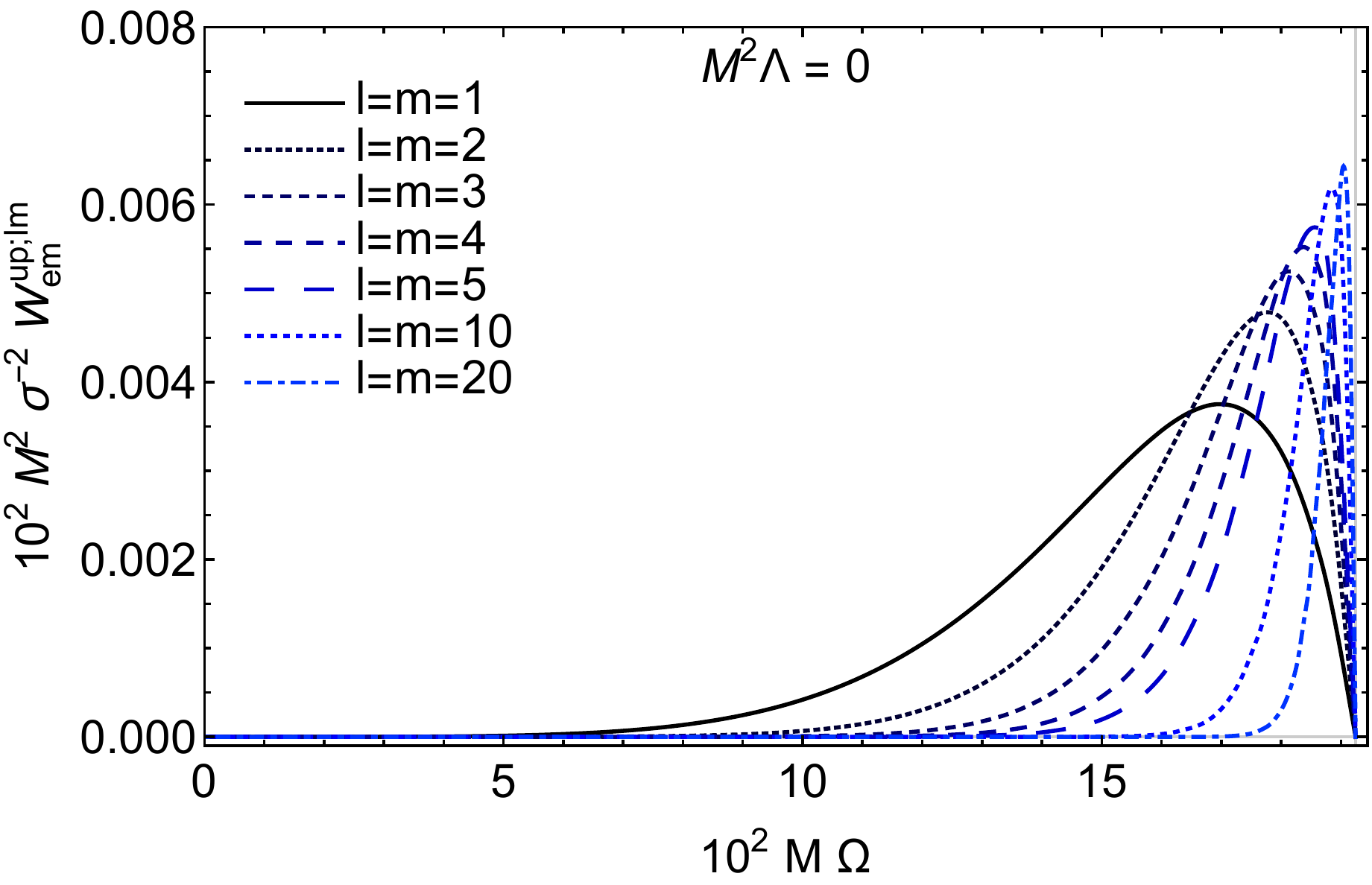} \vspace{0.3 cm}
\includegraphics[scale=0.45]{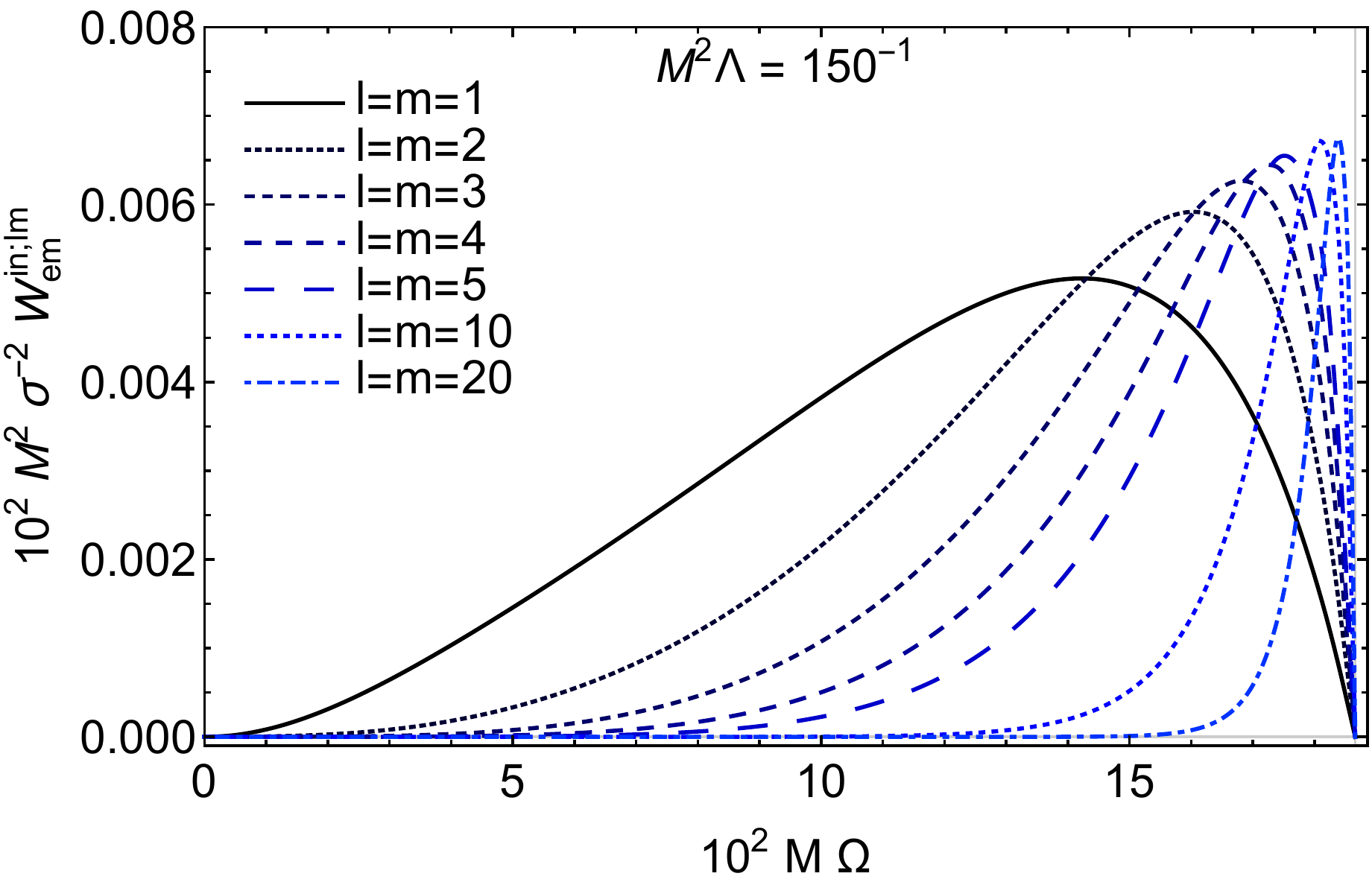} \hspace{0.2 cm}\vspace{0.3 cm}
\includegraphics[scale=0.45]{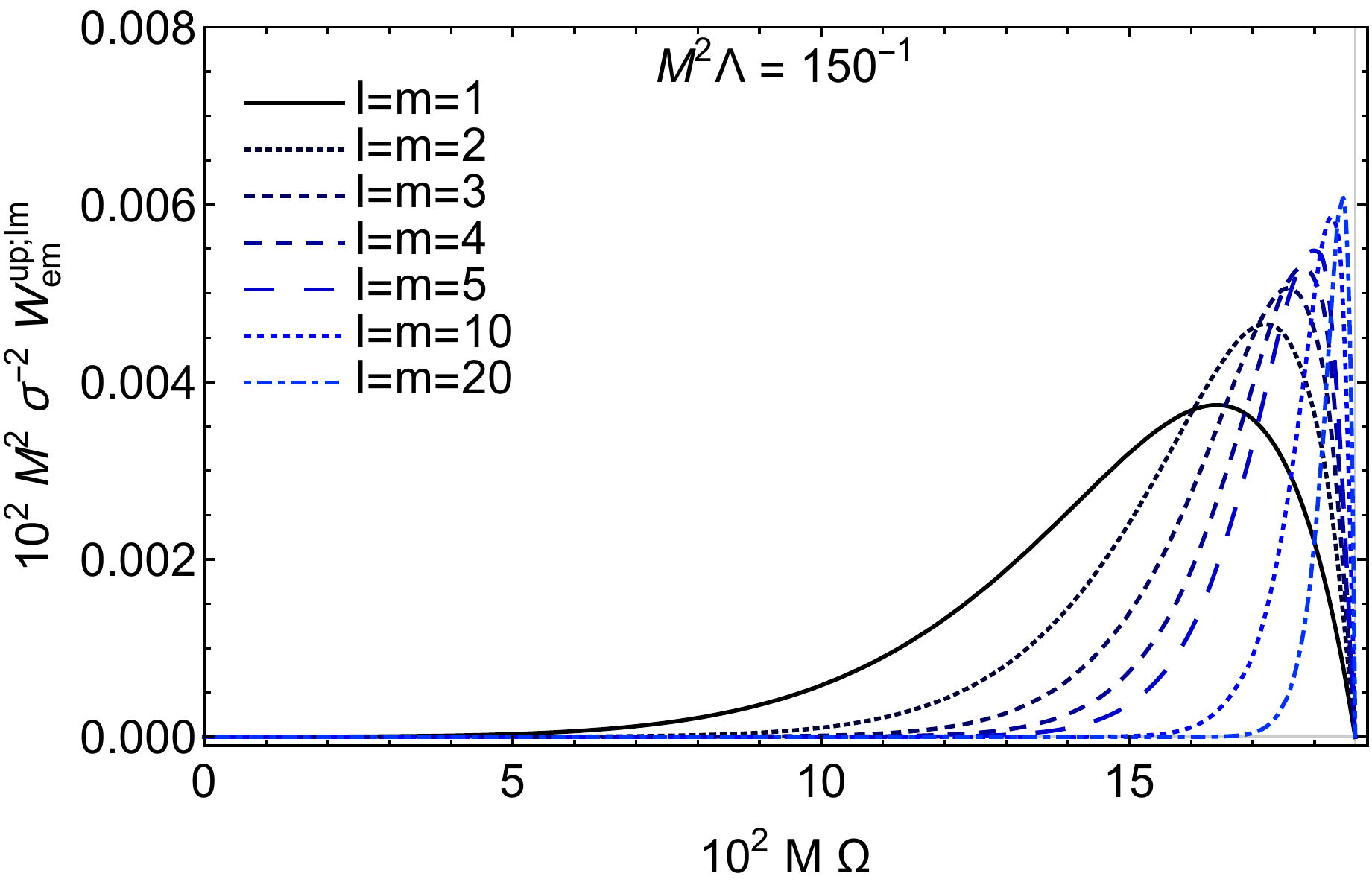} \vspace{0.3 cm}
\includegraphics[scale=0.45]{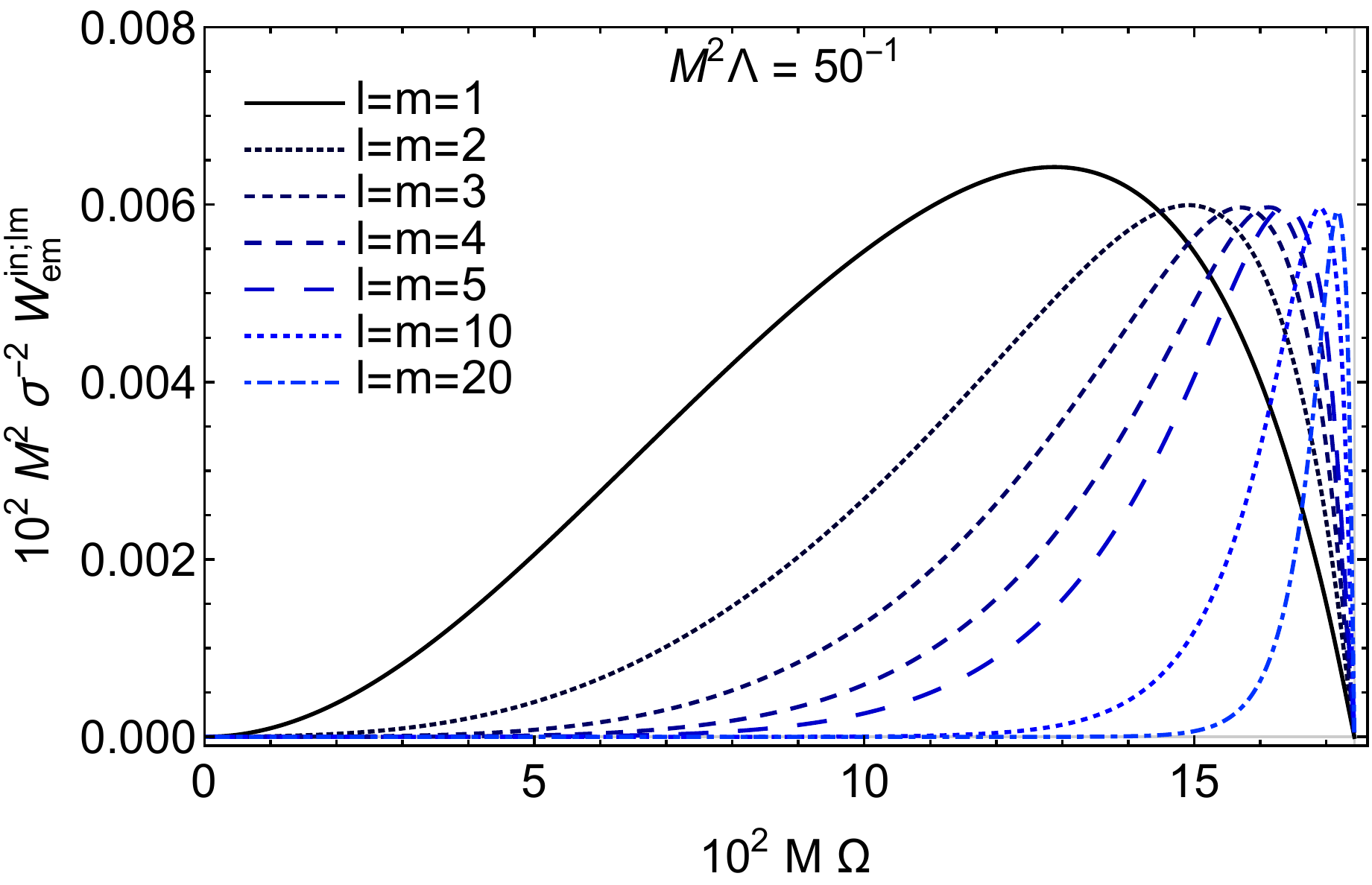} \hspace{0.2 cm}
\includegraphics[scale=0.45]{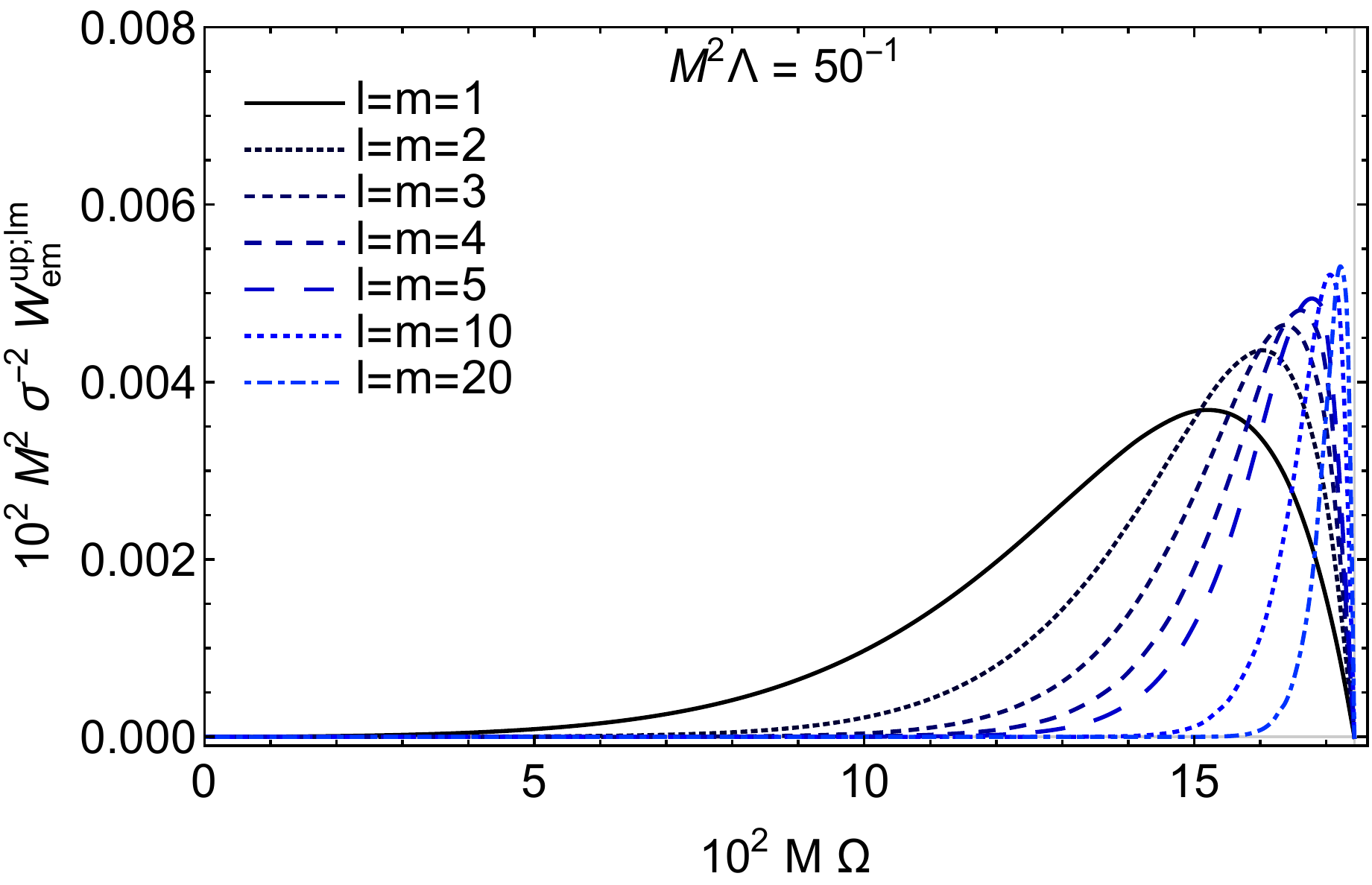} 
\includegraphics[scale=0.45]{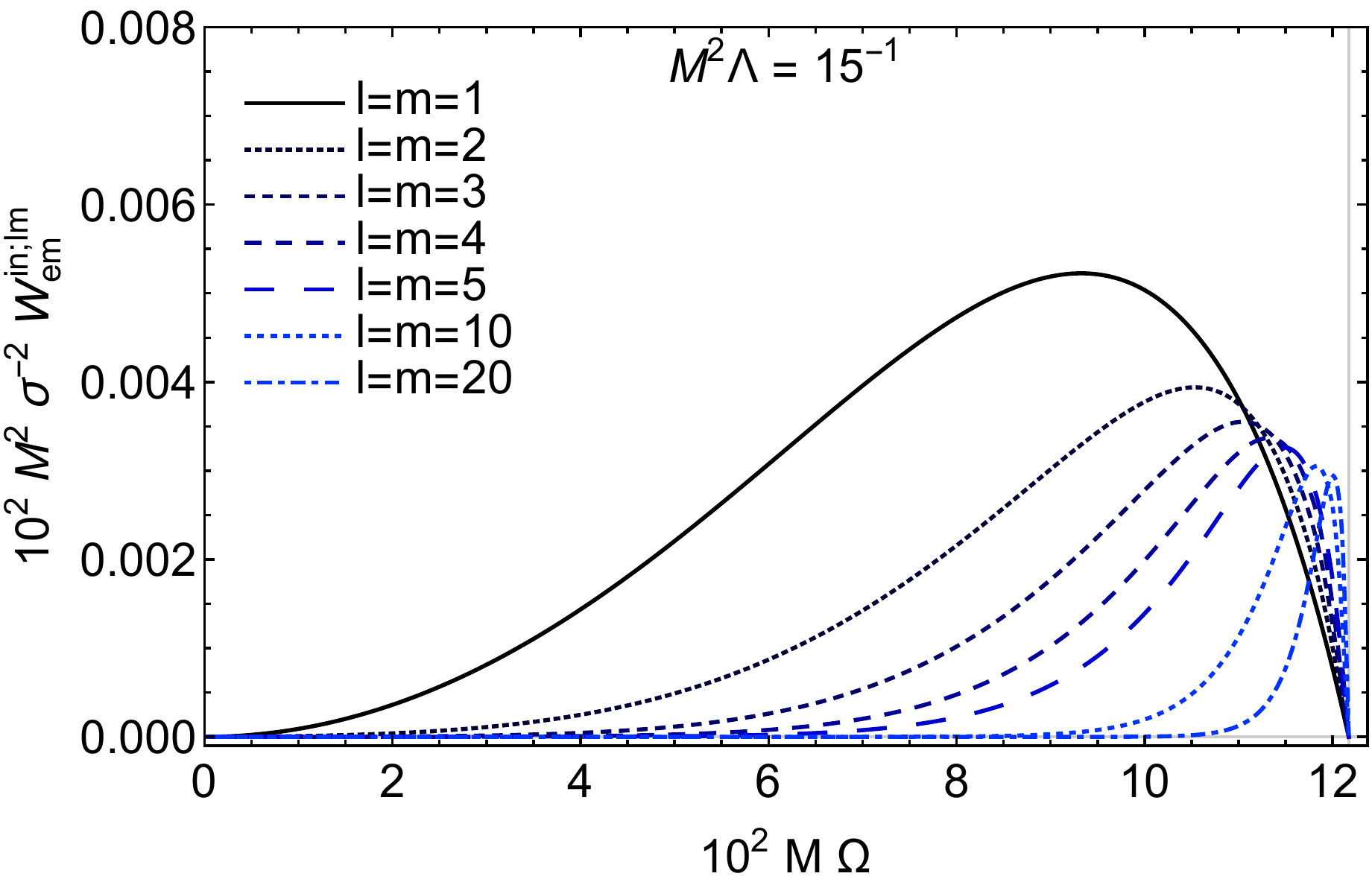} \hspace{0.2 cm}
\includegraphics[scale=0.45]{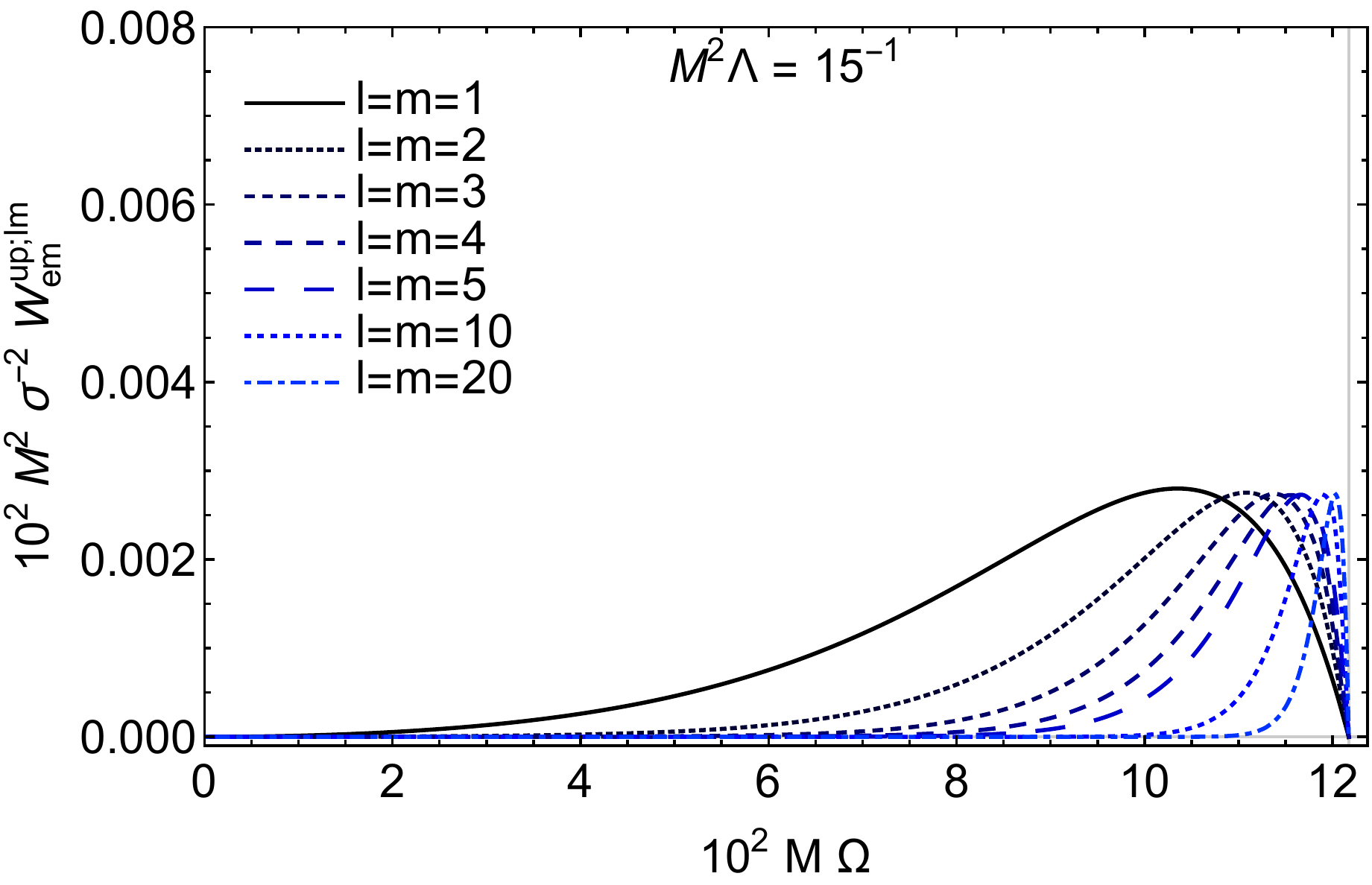} 
\caption{The emitted power, as a function of $\Omega,$ given by Eq.~(\ref{partial_power}), with different choices of $l=m,$ for $in$ (left) and $up$ (right) modes. We consider the black hole with different choices of the parameter $\Lambda,$ as indicated.}
\label{Pot_Parc_IN_e_UP_l_1_m_1}
\end{figure*}

In Fig.~\ref{Pot_Parc_IN_e_UP_l_1_m_1}, we plot separately the contribution from the $in$ and $up$ modes to the emitted power, for different choices of the cosmological constant $\Lambda$ and of the multipole numbers $l=m.$ We see that, when $M^{2}\Lambda < 150^{-1},$ the behavior of the emitted power, as a function of the angular velocity ($\Omega$) of the source is similar to that of the Schwarzschild case~\cite{crispino_2008}. On the other hand, for $M^2\Lambda > 150^{-1},$ we have an amplification of the power emitted by modes with lower values of $l=m.$ This effect is more evident for the $in$ modes, i.e., the modes purely incoming from the cosmological horizon ($H_c$).

\begin{figure*}
\center
\subfigure[]{\includegraphics[scale=0.45]{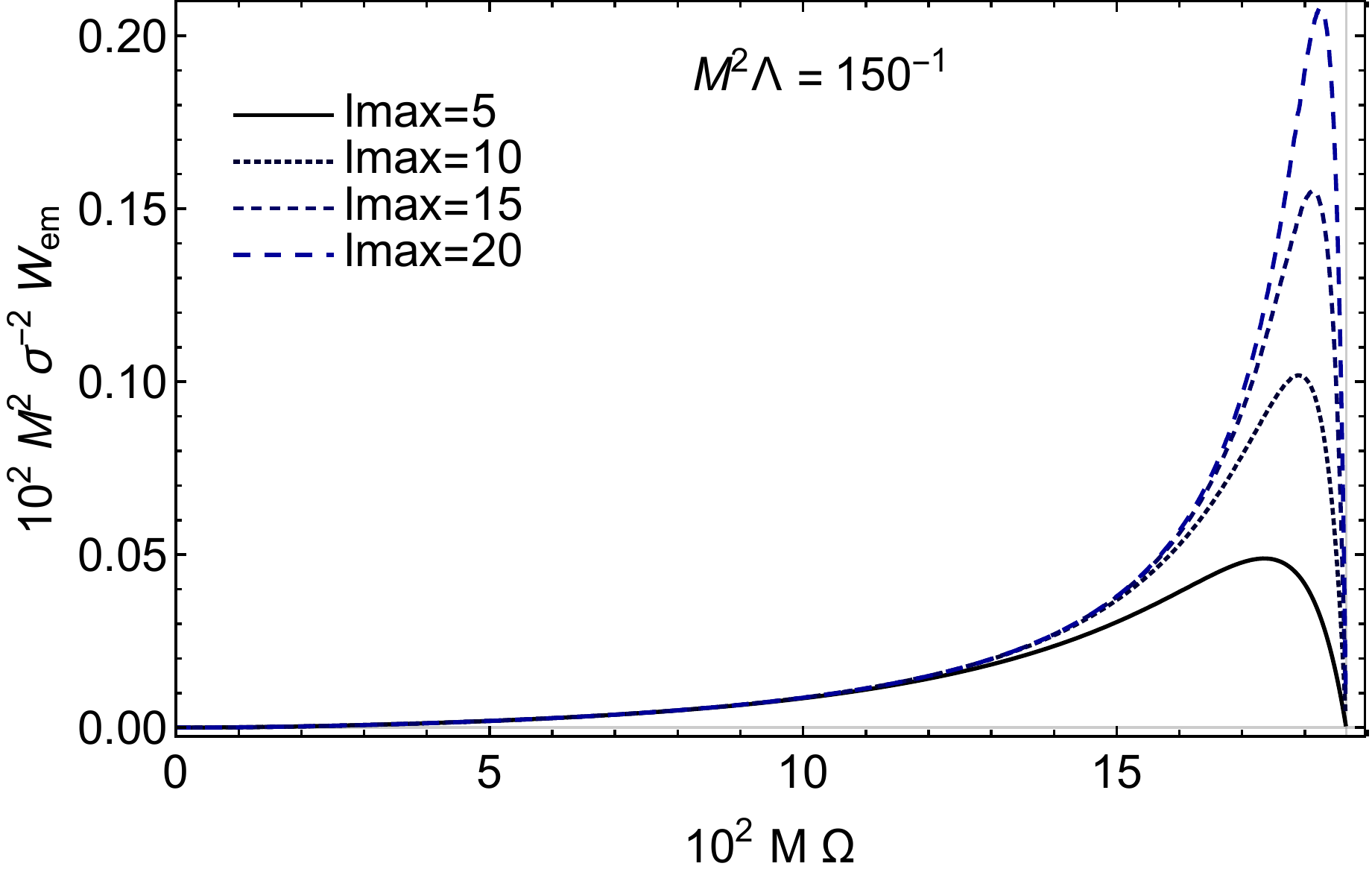}} \hspace{0.2 cm}
\subfigure[]{\includegraphics[scale=0.45]{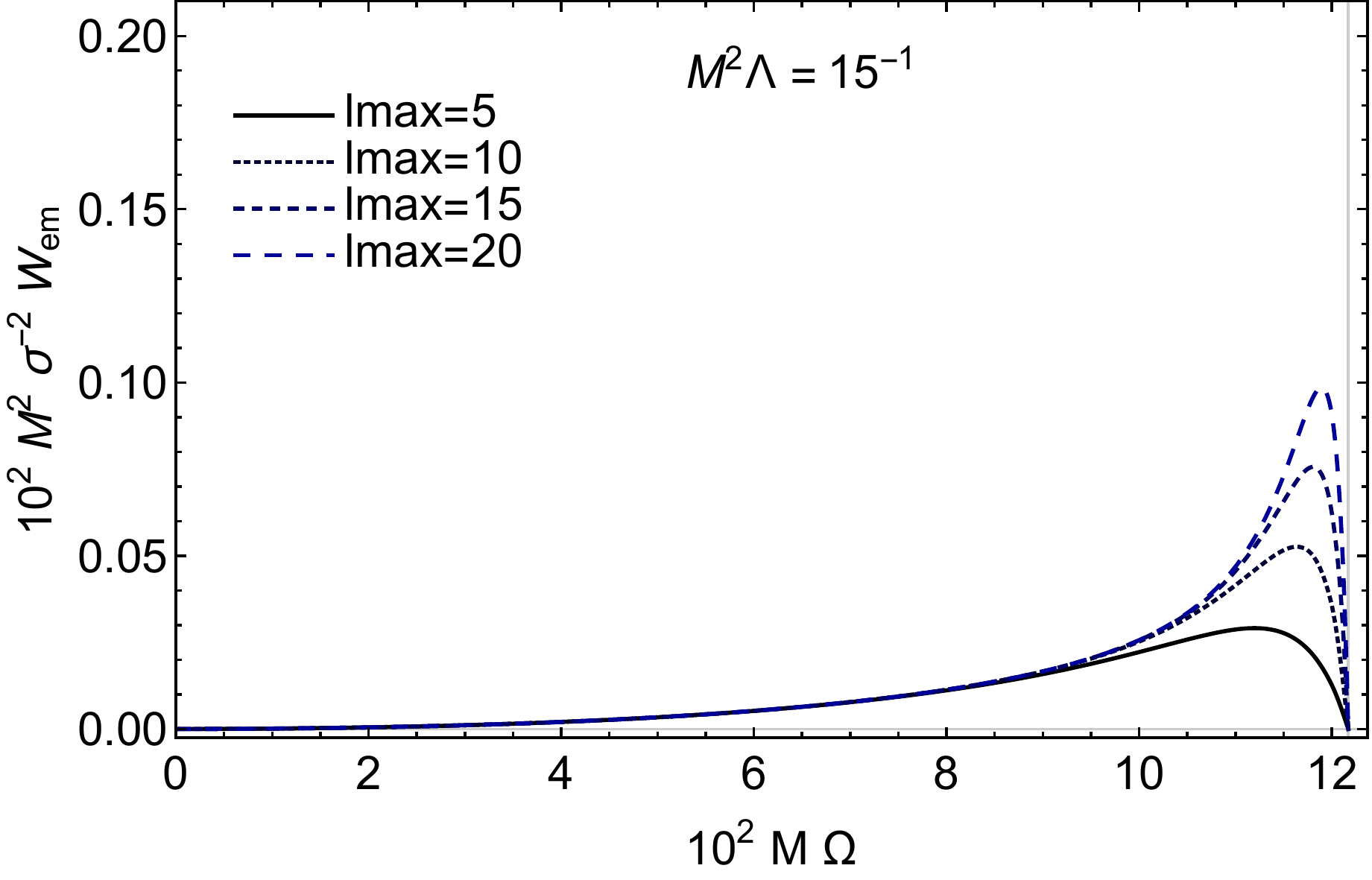}}
\caption{The total emitted power, given by Eq.~(\ref{total_power}), as a function of $\Omega,$ in the SdS spacetime with (a) $M^2\Lambda = 150^{-1}$ and (b) $M^2\Lambda = 15^{-1}$. The summations in $l$ were truncated at $l=l_{max},$ as indicated.}
\label{Pot_Tot_Lamb1_lmax}
\end{figure*}

In Fig.~\ref{Pot_Tot_Lamb1_lmax}, we plot the total emitted power, given by Eq.~(\ref{total_power}), for two choices of the cosmological constant. The $l$ summation was truncated at a maximum value $l=l_{max}.$ We see that, when the source lies relatively far from the black hole (where the $l=1$ mode contribution is dominant), the emitted power is basically the same for any choice of $l_{max},$ but as the orbit of the source approaches $r=3M,$ higher multipole modes start to contribute significantly, exhibiting a synchrotronic behavior of the emitted power. 

\section{Final Remarks}
\label{Sec_remarks}
In this paper we have used QFT in curved spacetime at tree level to investigate the scalar radiation emitted by a source orbiting a Schwarzschild--de Sitter (SdS) black hole. We have presented numerical results for the partial (with fixed $l$ and $m$) and total emitted powers, as functions of the angular velocity of the source.

We have found that the emitted power strongly depends on the value of the cosmological constant when $\Lambda > 150^{-1}M^{-2}.$ We have also shown that the emitted power associated to lower values of the multipole number $l$ is amplified as $\Lambda$ increases. In the Schwarzschild--anti--de Sitter geometry (for which $\Lambda < 0$), for sufficiently higher values of $\abs{\Lambda},$ an enhancement in the emitted power associated to higher values $l=m$ has been reported~\cite{cardoso_2002}. 

The scalar radiation considered in this paper have qualitative features similar to more realistic scenarios, as the ones of electromagnetic and gravitational radiation. Nevertheless, it is known that, in asymptotically flat spacetimes, the contribution of the high multipoles to the emitted power depends on the spin of the radiation field~\cite{Ruffini1972,Misner1974,bernar_2017,bernar_2018}. Thus, a similar investigation of fields with nonzero spin in asymptotically dS solutions, such as the SdS spacetime, will reveal the high multipoles behavior and their contribution to the emitted power, together with their relation to the black hole size.
It will also be interesting to extend this work to more general black hole spacetimes that are asymptotically dS, characterized by additional parameters, such as electric charge and angular momentum. 

\hspace{0.3cm}

\section*{Acknowledgments}
The authors would like to thank Coordena\c{c}\~ao de Aperfei\c{c}oamento de Pessoal de N\'ivel Superior (CAPES, Brazil) --- Finance Code 001, and Conselho Nacional de Desenvolvimento Cient\'ifico e Tecnol\'ogico (CNPq, Brazil) for partial financial support.
This research has also received funding from the European Union's Horizon 2020 research and innovation programme under the H2020-MSCA-RISE-2017 Grant No. FunFiCO-777740.



\begin{thebibliography}{99}
\bibitem{ligo1_2016}{B. P. Abbott \textit{et al.} (LIGO Scientific Collaboration and Virgo Collaboration), Observation of Gravitational Waves from a Binary Black Hole Merger, Phys. Rev. Lett.  \textbf{116}, 061102 (2016).}

\bibitem{ligo2_2016}{B. P. Abbott \textit{et al.} (LIGO Scientific Collaboration and Virgo Collaboration), GW151226: Observation of Gravitational Waves from a 22-Solar-Mass Binary Black Hole Coalescence, Phys. Rev. Lett.  \textbf{116}, 241103  (2016).}

\bibitem{EHT_sombra}{The Event Horizon Telescope Collaboration, First M87 Event Horizon Telescope Results. I. The Shadow of the Supermassive Black Hole, Astrophys. J. \textbf{875}, L1 (2019).}

\bibitem{wald_1984}{R. M. Wald, \textit{General Relativity} (The University of Chicago Press, Chicago,  1984).}

\bibitem{Bertietal2015}{E. Berti \textit{et al.}, Testing general relativity with present and future astrophysical observations, Class. Quantum Grav. \textbf{32}, 243001 (2015).}

\bibitem{Kiefer2005}C. Kiefer, Quantum gravity: General introduction and recent developments, Ann. Phys. (Amsterdam) \textbf{15}, 129 (2005).

\bibitem{birrel_1982}{N. D. Birrel, and P. C. W. Davies, \textit{Quantum fields in curved spacetime} (Cambridge University Press, Cambridge, 1982).}

\bibitem{parker_2009}{L. E. Parker, and D. J. Toms, \textit{Quantum Field Theory in Curved Spacetime: Quantized Fields and Gravity} (Cambridge University Press, Cambridge, 2009).}

\bibitem{Parker1969}{L. Parker, Quantized fields and particle creation in expanding universes. I, Phys. Rev. \textbf{183}, 1057 (1969).}

\bibitem{hawking_1975}{S. W. Hawking, Particle creation by black holes, Commun. Math. Phys. \textbf{43}, 199 (1975).}

\bibitem{Hawking1976}{S. W. Hawking, Breakdown of predictability in gravitational collapse, Phys. Rev. D \textbf{14}, 2460 (1976).}

\bibitem{Unruh1976}{W. Unruh, Notes on black-hole evaporation, Phys. Rev. D \textbf{14}, 870 (1976).}

\bibitem{Crispino2008}{L. C. B. Crispino, A. Higuchi, and G. E. A. Matsas, The Unruh effect and its applications, Rev. Mod. Phys. \textbf{80}, 787 (2008).}

\bibitem{misner_1972}{C. W. Misner, Interpretation of Gravitational-Wave Observations, Phys. Rev. Lett.  \textbf{28}, 994 (1972).}

\bibitem{misner_et_al_1972}{C. W. Misner, R. A. Breuer, D. R. Brill, P. L. Chrzanowski, H. G. Hughes, and C. M. Pereira,  Gravitational Synchrotron Radiation in the Schwarzschild Geometry, Phys. Rev. Lett.  \textbf{28}, 998 (1972).}

\bibitem{crispino_2000}{L. C. B. Crispino, A. Higuchi, and G. E. A. Matsas, Scalar radiation emitted from a source rotating around a black hole, Class. Quantum Grav. \textbf{17}, 19 (2000); Corrigendum \textbf{33}, 209502 (2016).}

\bibitem{castineiras_2007}{J. Casti\~neiras, L. C. B. Crispino, and D. P. M. Filho, Source coupled to the massive scalar field orbiting a stellar object, Phys. Rev. D \textbf{75}, 024012 (2007).}

\bibitem{crispino_2008}{L. C. B. Crispino, Synchrotron scalar radiation from a source in ultrarelativistic circular orbits around a Schwarzschild black hole, Phys. Rev. D \textbf{77}, 047503 (2008).}

\bibitem{crispino_2009}{L. C. B. Crispino, A. R. R. da Silva, and G. E. A. Matsas, Scalar radiation emitted from a rotating source around a Reissner-Nordstr\"{o}m black hole, Phys. Rev. D \textbf{79}, 024004 (2009).}

\bibitem{macedo_2012}{C. F. B. Macedo, L. C. B. Crispino, and V. Cardoso, Semiclassical analysis of the scalar geodesic synchrotron radiation in Kerr spacetime, Phys. Rev. D \textbf{86}, 024002 (2012).}

\bibitem{bernar_2019}{R. P. Bernar, and L. C. B. Crispino, Scalar radiation from a source rotating around a regular black hole, Phys. Rev. D \textbf{100}, 024012 (2019).}

\bibitem{castineiras_2005}{J. Casti\~neiras, L. C. B. Crispino, R. Murta, and G. E. A. Matsas, Semiclassical approach to black hole absorption of electromagnetic radiation emitted by a rotating charge, Phys. Rev. D \textbf{71}, 104013 (2005).}

\bibitem{bernar_2017}{R. P. Bernar, L. C. B. Crispino, and A. Higuchi, Gravitational waves emitted by a particle rotating around a Schwarzschild black hole: A semiclassical approach, Phys. Rev. D \textbf{95}, 064042 (2017).}

\bibitem{bernar_2018}{R. P. Bernar, L. C. B. Crispino, and A. Higuchi, Circular geodesic radiation in Schwarzschild spacetime: A semiclassical approach, Int. J. Mod. Phys. D \textbf{27}, 1843002 (2018).}

\bibitem{cardoso_2002}{V. Cardoso, and J. P. Lemos, Scalar synchrotron radiation in the Schwarzschild-anti-de Sitter geometry, Phys. Rev. D \textbf{65}, 104033 (2002).}

\bibitem{desitter_1917_1}{W. de Sitter, On the relativity of inertia: Remarks concerning Einstein's latest hypothesis, Proc. Kon. Ned. Akad. Wet. \textbf{19}, 1217 (1917).}

\bibitem{desitter_1917_2}{W. de Sitter, On the curvature of space, Proc. Kon. Ned. Akad. Wet. \textbf{20}, 229 (1917).}

\bibitem{hawking_1973}{S. W. Hawking, and G. F. R. Ellis, \textit{The large scale structure of space-time} (Cambridge University Press, Cambridge, 1973).}

\bibitem{schrodinger}{E. Schr\"{o}dinger, \textit{Expanding universes} (Cambridge University Press, Cambridge, 1956).}

\bibitem{riess_1998}{A. G. Riess \textit{et al.}, Observational Evidence from Supernovae for an Accelerating Universe and a Cosmological Constant, Astron. J. \textbf{116}, 1009 (1998).}

\bibitem{perlmutter_1999}{S. Perlmutter \textit{et al.}, Measurements of $\Omega$ and $\Lambda$ from 42 high-redshift supernovae, Astrophys. J. \textbf{517}, 565 (1999).}

\bibitem{kottler_1918}{F. Kottler, \"{U}ber die physikalischen grundlagen der Einsteinschen gravitationstheorie, Ann. Phys. (N.Y.) \textbf{361}, 401 (1918).}

\bibitem{stuchlik_1999}{Z. Stuchl\'ik, and S. Hled\'ik, Some properties of the Schwarzschild--de Sitter and Schwarzschild-anti-de Sitter spacetimes, Phys. Rev. D \textbf{60}, 044006 (1999).}

\bibitem{akcay_2011}{S. Akcay, and R. Matzner, The Kerr-de Sitter universe, Class. Quantum Grav. \textbf{28}, 085012 (2011).}

\bibitem{rindler_2006}{W. Rindler, \textit{Relativity: Special, General, and Cosmological} (Oxford University Press, New York, 2006).}

\bibitem{howes_1979}{ R. J. Howes, Existence and stability of circular orbits in a Schwarzschild field with nonvanishing cosmological constant, Aust. J. Phys. \textbf{32}, 293 (1979).}

\bibitem{boonserm_2019}{P. Boonserm, T. Ngampitipan, A. Simpson, and M. Visser, Innermost and outermost stable circular orbits in the presence of positive cosmological constant, Phys. Rev. D \textbf{101}, 024050 (2020).}

\bibitem{boulware_1975}{D. G. Boulware, Quantum field theory in Schwarzschild and Rindler spaces, Phys. Rev. D \textbf{11}, 1404 (1975).}

\bibitem{higuchi_1987}{A. Higuchi, Quantisation of scalar and vector fields inside the cosmological event horizon and its application to the Hawking effect, Class. Quantum Grav. \textbf{4}, 721 (1987).}

\bibitem{ashtekar_1975}{A. Ashtekar, and A. Magnon, Quantum fields in curved space-times, Proc. R. Soc. London A. \textbf{346}, 375 (1975).}

\bibitem{fulling_1973}{S. A. Fulling, Nonuniqueness of canonical field quantization in Riemannian space-time, Phys. Rev. D \textbf{7}, 2850 (1973).}

\bibitem{itzykson_1980}{C. Itzykson, and J.-B. Zuber, \textit{Quantum Field Theory} (McGraw-Hill Inc., New York, 1980).}

\bibitem{breuer_1975}{R. A. Breuer, \textit{Gravitational perturbation theory and Synchrotron Radiation}, \textit{Lecture Notes in physics Vol. 44} (Springer-Verlag, Heidelberg, 1975).}

\bibitem{crispino_1998}{L. C. B. Crispino, A. Higuchi, and G. E. A. Matsas, Interaction of Hawking radiation and a static electric charge, Phys. Rev. D \textbf{58}, 084027 (1998).}

\bibitem{gradshteyn_1980}{I. S. Gradshteyn, and I. M. Ryzhik, \textit{Table of Integrals, Series and Products, Corrected and Enlarged Edition} (Academic Press, New York, 1980).}

\bibitem{Ruffini1972}{M. Davis, R. Ruffini, J. Tiomno, and F. Zerilli, Can Synchrotron Gravitational Radiation Exist?, Phys. Rev. Lett. \textbf{28}, 1352 (1972).}

\bibitem{breuer_1973}{R. A. Breuer, R. Ruffini, J. Tiomno, and C. V. Vishveshwara, Vector and Tensor Radiation from Schwarzschild Relativistic Circular Geodesics, Phys. Rev. D \textbf{7}, 1002 (1973).}

\bibitem{Misner1974}{P. L. Chrzanowski, and C. W. Misner, Geodesic synchrotron radiation in the Kerr geometry by the method of asymptotically factorized Green's functions, Phys. Rev. D \textbf{10}, 1701 (1974).}


\end{thebibliography}
\end{document}